\documentclass[aps,
pra,
		reprint,
        superscriptaddress,
        shortbibliography,
		  nofootinbib,
		floatfix,
		notitlepage, twocolumn
		]{revtex4-2}

\usepackage{fullpage}
\usepackage{graphicx}  
\usepackage{amsmath}   
\usepackage{hyperref}
\hypersetup{colorlinks=true,linkcolor=blue}
\usepackage{amssymb}   
\usepackage{bm} 
\usepackage{dcolumn}
\usepackage{color}
\usepackage{xcolor}
\usepackage{mathrsfs}
\usepackage{varioref}
\usepackage{braket}
\usepackage{comment}
\usepackage{tikz}
\usepackage{subcaption}
\usepackage{ragged2e}
\usepackage[font=small,labelfont=bf,
   justification=justified,
   format=plain]{caption}
\usepackage{lipsum}
\usepackage{empheq}

\usetikzlibrary{positioning}
\usepackage{orcidlink}

\newcommand{\Sig}{\mathbf{\Sigma}}
\newcommand{\J}{\mathbf{\mathcal{J}}}
\newcommand{\A}{\mathbf{\mathcal{A}}}
\labelformat{section}{Section #1} 
\labelformat{subsection}{Section #1} 
\labelformat{subsubsection}{Section #1}
\labelformat{subsubsubsection}{Section #1}
\labelformat{equation}{Eq.~(#1)} 
\labelformat{figure}{Fig.~#1} 
\labelformat{subfigure}{Fig.~\thefigure#1} 
\labelformat{table}{Tab.~#1} 
\labelformat{appendix}{Appendix #1}

\begin{document}
\title{Thermalization from quenching in coupled oscillators}
\author{M Harinarayanan\orcidlink{0009-0008-4105-1441}}
\email{mharinarayanan7@gmail.com}
\affiliation{Bharata Mata College, Thrikkakara, Kochi 682021, Kerala, India}
\affiliation{Department of Physics, Indian Institute of Technology Madras, Chennai 600036, India}
\author{Karthik Rajeev\orcidlink{0000-0003-3193-1900}}
\email{karthik.rajeev@ed.ac.uk}
\affiliation{Higgs Centre, School of Physics and Astronomy, University of Edinburgh, EH9 3FD, UK}

\begin{abstract}
  We introduce a finite-time protocol that thermalizes a quantum harmonic oscillator, initially in its ground state, without requiring a macroscopic bath. The method uses a second oscillator as an effective environment and implements sudden quenches of the oscillator frequencies and coupling. Owing to the Gaussian nature of the dynamics, the thermalization condition reduces to three solvable equations, yielding exact analytic solutions for a dense discrete set of temperatures and numerical solutions in all other cases. Any target temperature can be approximated with arbitrary precision, with a trade-off between speed and accuracy. The simplicity of the protocol makes it a promising tool for rapid, controlled thermalization in quantum thermodynamics experiments and state preparation.
\end{abstract}

\maketitle

\section{Introduction}\label{sec:intro}
The concept of thermalization is central to both classical and quantum physics, underpinning the emergence of equilibrium properties from microscopic dynamics. Thermality also remains a key topic in contemporary research across quantum information\cite{Goold2016role,deffner2019qtqi}, many-body physics\cite{Millen_Xuereb_2016}, and even quantum gravity\cite{Witten:2024upt,Harlow:2014yka}.

 Addressing this aspect for quantum systems is the focus of the rapidly growing field of quantum thermodynamics (for recent reviews, see \cite{Vinjanampathy_2016,Myers:2022lvm}). In addition to its fundamental significance, advancing our understanding of quantum thermodynamic processes is expected to play an important role in the development of future technologies, particularly at the nanoscale\cite{H_nggi_2009,Horodecki_2013}, where quantum effects become dominant.  In quantum simulations\cite{Georgescu:2013oza}---positioned at the intersection of advanced technology and fundamental physics---efficient preparation of thermal states plays a crucial role in multiple applications. Therefore, there are both theoretical and experimental efforts to improve and optimize the preparation of thermal states of quantum systems\cite{Chen:2023cuc,ding2025therm}. 

 The quantum harmonic oscillator (QHO) is not only a fundamental exactly solvable model in quantum mechanics but also a key component of various physical platforms—from vibrational modes in trapped ions to superconducting resonators in circuit QED, as well as optomechanical mirrors and nanomechanical membranes \cite{Wineland:1997mg,Wallraff2004,Aspelmeyer:2013lha,Bachtold:2022hnr}. Consequently, the QHO holds a central role in the study of thermodynamics of quantum systems\cite{Kosloff_2017,Deffner_2018,Serafini_2020,Singh_2020}.

Traditionally, thermalization of a quantum harmonic oscillator (QHO) is achieved by coupling it to a heat bath and allowing equilibration over long timescales. Recent works have extended this framework to finite baths \cite{PhysRevA.91.020502,PhysRevE.93.062106,Reid_2017} and developed formulations that treat the bath and system on equal footing \cite{Bera_2019}. In parallel, advances in nonequilibrium quantum control have motivated alternative approaches, including engineered \textit{shortcuts} to inherently slow processes such as adiabatic evolution \cite{Torrontegui:2013sge,del_Campo_2019}. Finite-time thermalization has been approached through several distinct 
methods. Shortcuts to adiabaticity (STA) extended to open quantum 
systems~\cite{Dupays:2020sta, Alipour:2020sad, Santos:2024sta} achieve 
rapid thermalization through engineered dephasing channels or 
counterdiabatic driving fields derived from the instantaneous eigenstates 
of the system. Engineered reservoir approaches~\cite{Poyatos:1996qre, Kienzler_2015, So:2025etr, Metcalf:2020etc} drive the system to a 
steady state determined by externally imposed bath parameters.

Building on these ideas, our work proposes a protocol that achieves \textit{exact} thermalization of QHO within a \textit{finite time} by substituting the conventional heat bath with a second \textit{identical} QHO and time-dependent control. Thermalization is driven by a sequence of sudden quenches applied to the frequencies and couplings of the two-oscillator system, effectively preparing the target thermal state without the need for long interactions with a macroscopic reservoir. 

While this system is already of interest for probing and challenging conventional notions of thermodynamics---the finite-time nature questions the asymptotic assumption, while the quench dynamics challenge adiabaticity--- we present our analytical model as a step toward a deeper question: can thermodynamic experiments be realized with a single ion acting simultaneously as both system and bath? The underlying intuition is that a two-oscillator system can be physically implemented using the two transverse motional degrees of freedom of an ion in an effective 2D trap. The details of this experimental realization will be addressed in future work.

The paper is structured as follows: We begin by briefly reviewing relevant aspects of pure Gaussian states of a system of two bosonic oscillators, and setting up notations and conventions in \ref{sec:gaussian}. Following that, in \ref{sec:setup}, we introduce the general set-up on which our prescription is based. In \ref{sec:main}, we outline the main result of this paper, namely, the prescription to achieve thermalization of a bosonic oscillator in \textit{finite time} using our setup. Finally, we conclude with a summary and future outlook in \ref{sec:discussion}. ($\hbar=k_{B}=1$, unless otherwise stated.)

\section{Review of two-mode Gaussian pure states}\label{sec:gaussian}

The core idea behind the thermal state preparation method proposed in this work relies on the properties of Gaussian pure states associated with a two-oscillator system. To set the stage, we begin with a brief review of the essential concepts and establish the notation that will be used throughout the rest of the paper.

 We label the two oscillators as \textit{oscillator}-1 (the system) and \textit{oscillator}-2 (the environment), with coordinates $x_1$ and $x_2$, respectively. For simplicity of the discussion, the oscillators have identical mass  $m$ and frequency $\omega$. A general \textit{undisplaced} Gaussian pure state of the combined system can be described by the wavefunction:
\begin{align}\label{Eq:gaussian_wf}
	\psi(x_1,x_2)=\mathcal{N}\exp\left[-\sum_{i,j=1}^{2}\frac{A_{ij}}{2}x_ix_j\right],
\end{align}
where $\mathcal{N}$ is the normalization factor and the complex coefficients $A_{ij}$ satisfy $\textrm{Re}[A_{11}],\textrm{Re}[A_{22}]>0$ and $2\textrm{Re}[A_{11}]\textrm{Re}[A_{22}]>\textrm{Re}[A_{12}^2]$.  

Since \textit{oscillator}-1 is the system we want to do manipulations on, we shall be interested in the reduced density matrix $\rho^{(1)}_{x_1\,x_1'}$ of the same, obtained by tracing over $x_2$ and which takes the form:
\begin{align}
	\rho^{\rm (1)}_{x_1'\,x_1}&=|\mathcal{N}|^2\int_{-\infty}^{\infty}\psi^*(x_1',x_2)\psi(x_1,x_2) dx_2\,,\\\nonumber
 &=|\mathcal{N}|^2e^{-\frac{m\omega}{2}\left[(X+i Y)x_1'{}^2+(X-i Y)x_1^2+2Zx_1'x_1\right]}\,.
\end{align}
Here, the real dimensionless parameters $X,Y$ and $Z$ are defined by
\begin{align}\label{eq:def_R}
    X+iY&=\frac{1}{m\omega}\left(A_{11}-\frac{A_{12}^2}{2\textrm{Re}[A_{22}]}\right)\,,\\
    Z&=- \frac{1}{m \omega}\left(\frac{|A_{12}|^2}{2\textrm{Re}[A_{22}]}\right)\,.
\end{align}
These parameters are, of course, related to the components of the covariance matrix through
\begin{align}\label{eq:def_sigma}
    \begin{bmatrix}
        \Sigma_{xx}&\Sigma_{xp}\\
        \Sigma_{px}&\Sigma_{pp}
    \end{bmatrix}
    =\frac{1}{2}
    \begin{bmatrix}
        \frac{1}{X+Z}&\frac{i(X+iY +Z)}{X+Z}\\
        \frac{-i(X-iY +Z)}{X+Z}&\frac{X^2+Y^2-Z^2}{X+Z}
    \end{bmatrix}\,.
\end{align}
The reduced density matrix is therefore equally fully characterized by the real variables $X,Y$ and $Z$. 
Since we are interested in dynamical density matrices, these parameters vary with time, allowing the evolution of the density matrix to be conveniently represented by the three-dimensional curve
\begin{align}
    \vec{R}(t)\equiv(X(t),Y(t),Z(t))\,.
\end{align}

\begin{figure}[t]
    \includegraphics[width=0.75\linewidth]{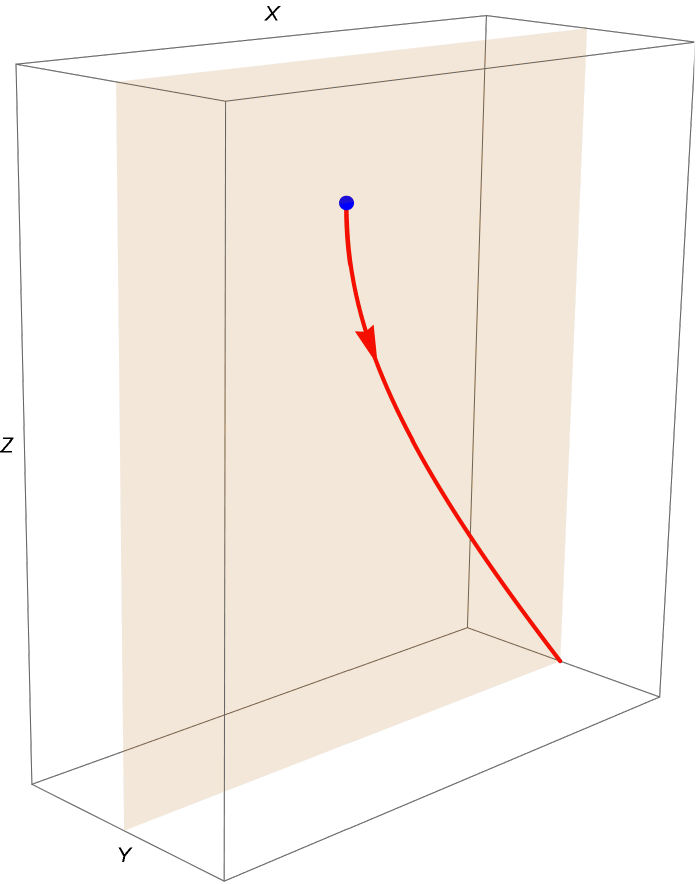}
    \caption{
        \justifying The family of thermal density matrices is represented by the curve $\vec{R}_\beta = \left( \coth(\beta\omega),\, 0,\ -\mathrm{cosech}(\beta\omega) \right)$. The curve lies entirely in the $X$--$Z$ plane, shown as the shaded region, where it is described by part of the hyperbola $Z=-\sqrt{X^2-1}$. The blue dot indicates the ground state, and the arrow shows the direction of increasing temperature.}
    \label{fig:therm_curve}
\end{figure}

The main goal of this paper is to prescribe a method to dynamically generate a thermal state, for which the density matrix $\rho_{\beta}$ takes the form  
\begin{align}
    \braket{x_1|\rho_{\beta}|x_1'}\propto e^{-\frac{m \omega}{2}\left[(x_1^2+x_1'{}^2)\coth{\beta\omega}-2x_1x_1'\textrm{csch}(\beta\omega)\right]}\,,
\end{align}
where $\beta=1/T$ is the inverse temperature. This translates to 
\begin{align}\label{eq:R_beta}
    \vec{R}\rightarrow \vec{R}_{\beta}\equiv \left(\coth(\beta\omega),0,-\textrm{cosech}(\beta\omega)\right)\,,
\end{align}
which describes a one-parameter curve in the $\vec{R}$-space as shown in \ref{fig:therm_curve}.
Since the above equations impose three distinct constraints, it is natural to expect that a suitably designed Hamiltonian evolution for the coupled oscillators---with three tunable parameters---could dynamically produce a desired thermal state, given appropriate adjustments of those parameters. In the following section, we demonstrate that this is indeed achievable through a concrete example.

\section{The set up}\label{sec:setup}

\begin{figure}[t]
    \centering
    \begin{subfigure}[t]{0.5\textwidth}
        \centering
        \includegraphics[width=.65\textwidth]{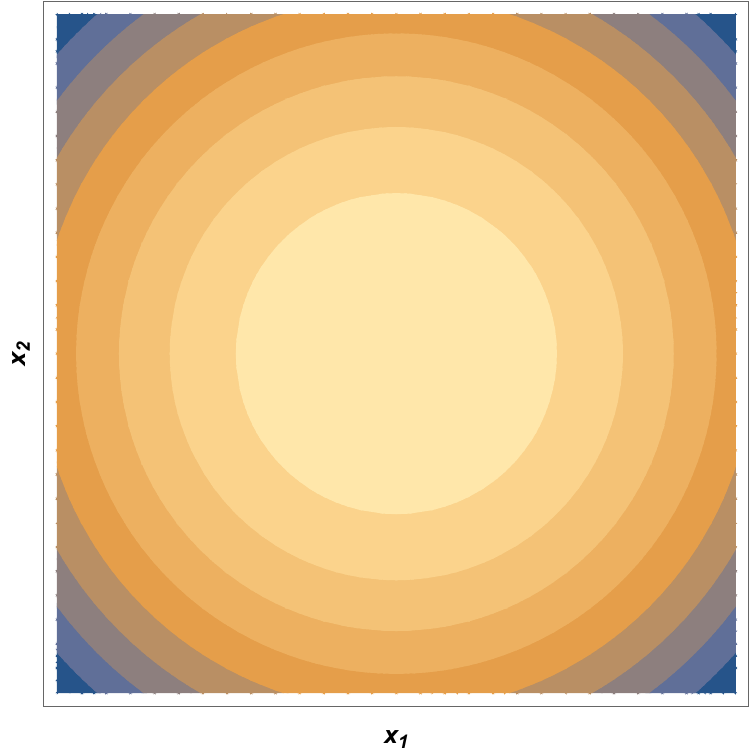}
       \caption{}
    \end{subfigure}%
    \\
    \begin{subfigure}[t]{0.5\textwidth}
        \centering
        \includegraphics[width=.65\textwidth]{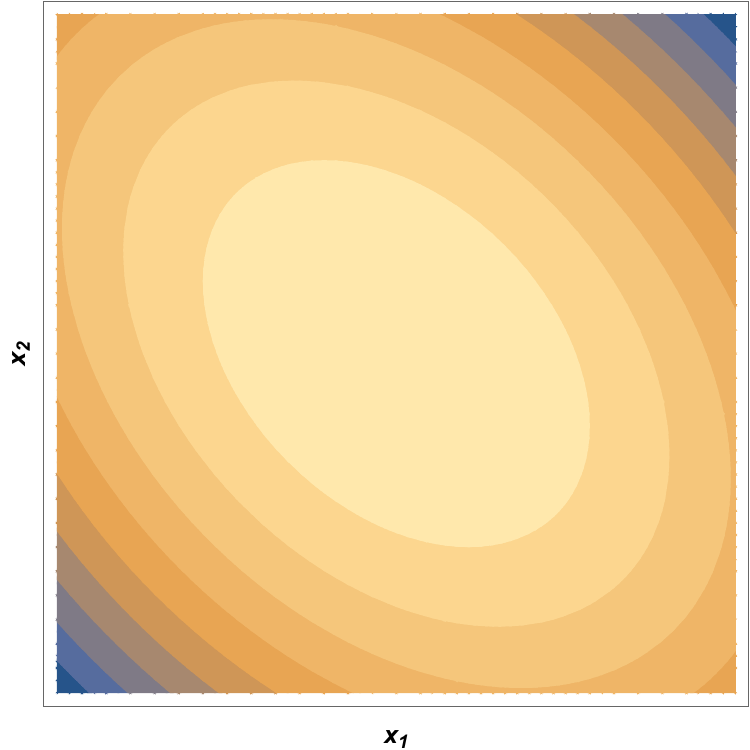}
        \caption{}
    \end{subfigure}
    \caption{\justifying Contour representation of the effective 2D potential of the coupled oscillator system, for typical values of $\omega'/\omega$ and $k/\omega^2$, when the system is in \textbf{(a)} the uncoupled phase, and \textbf{(b)} the coupled phase ($0<t<\tau$).}\label{figure:V12}
\end{figure}

We consider a system of two coupled oscillators described by a Hamiltonian of the form:
\begin{align}\label{Eq:hamiltonian}
	H=\frac{1}{2m}p_1^2+\frac{1}{2m}p_2^2&+\frac{1}{2}m\Omega^2(t)x_1^2+\frac{1}{2}m\Omega^2(t)x_2^2\\\nonumber
    &+\frac{1}{2}\mathcal{K}(t)(x_1-x_2)^2\,,
\end{align}
where we shall refer to $\mathcal{K}$ as the `coupling'. Henceforth, without loss of generality, we shall set $m=1$. The above Hamiltonian can also be interpreted as describing a particle confined in a time-dependent two-dimensional harmonic potential well. For our proposal, we assume the following time dependence for the frequency and coupling:
\begin{align}\label{eq:defin_quench}
	\mathcal{K}(t)&=\begin{cases}
	k\quad &;\quad 0<t<\tau\\
	0\quad &;\quad \textrm{otherwise}
	\end{cases}\,,\\\nonumber
	\Omega(t)&=\begin{cases}
	\omega\quad &;\quad t<0\textrm{ and }t>\tau\\
	\omega'\quad &;\quad 0<t<\tau 
	\end{cases}\,,
\end{align}
where, $k$, $\omega'$ and $\tau$ are tuneable. The time dependence of the potential, when realised in the context of trapped ions(such as in \cite{Wineland:1997mg}), can then be visualized as a kind of squeezing of the equipotential contours as shown in \ref{figure:V12}.

As alluded to before, a potential experimental realization of the above setup 
would be a single ion in a Paul trap\cite{Leibfried2003, Wineland:1997mg}, where the two transverse radial motional modes serve as the oscillators, with one mode 
acting as the system and the other as the environment. In this context, experimental control could be achieved by modulating a set of external RF quadrupole potentials~\cite{Ding:2017qpo}, which governs both the frequency quenches and the inter-mode coupling. Moreover, the ground 
state for the system can be prepared by laser cooling the motional degree of 
freedom~\cite{Diedrich1989,monroe1995resolved}. Primary sources of experimental error include fluctuations in trapping potentials causing instabilities in normal mode 
frequencies~\cite{Turchette2000}, timing noise in the quench sequences affecting the precision of $\tau$, and environmental factors such as ambient heating and off-resonant scattering~\cite{Brownnutt2015,Wineland:1997mg}, which impose a lower bound on achievable 
temperatures. Here, however, our focus is on developing the analytical aspects of the proposed protocol, while the details of its experimental realization will be addressed elsewhere.

We now return to the quantum dynamics of time-dependent coupled oscillators, which is a well-understood problem. In the special case of Gaussian pure states, the dynamics can be fully described in terms of solutions to the classical equations of motion of the normal modes. In terms of the normal modes--- which are $x_{\pm}=(x_1\pm x_2)/\sqrt{2}$, with normal mode frequencies $\Omega_{+}(t)=\Omega(t)$ and $\Omega_{-}(t)=\sqrt{\Omega^2(t)+2\mathcal{K}(t)}$--- the problem reduces to that of two time dependent uncoupled oscillators. This standard procedure leads to a Gaussian wavefunction of the form in \ref{Eq:gaussian_wf}, with $A_{ij}$ being functions of time. Specifically, \vspace{-.4cm}
\begin{align}\label{eq:def_A11}
    A_{11}(t)=A_{22}(t)=\sum_{\sigma=+,-}\left(\frac{\Omega_{\sigma}(0)}{2b_{\sigma}^2}-\frac{i \dot{b}_{\sigma}}{2b_{\sigma}}\right)\,,\\\label{eq:def_A12}
    A_{12}(t)=\sum_{\sigma=+,-}\sigma\left(\frac{\Omega_{\sigma}(0)}{2b_{\sigma}^2}-\frac{i \dot{b}_{\sigma}}{2b_{\sigma}}\right)\,,
\end{align}
where, the functions $b_{\pm}(t)$ satisfy 
\begin{align}\label{eq:ermakov}	\ddot{b}_{\pm}+\Omega_{\pm}^2(t)b_{\pm}=\frac{\Omega_{\pm}^2(0)}{b_{\pm}^3}\,.
\end{align}
The above \textit{non-linear} equations are nothing but the Ermakov equations \cite{ermakov1880,lewis1967classical,lewis1968class} corresponding to the normal modes.

We now focus on the specific quench-type time dependence defined in Equation \ref{eq:defin_quench}, for which $\Omega_{+}(0)=\omega'$ and $\Omega_{-}(0)=\sqrt{\omega'{}^2+2k}$. We demand that the full system is initialized in the ground state at $t=0$, when the system is uncoupled. This translates to the initial conditions of the form $b_{\pm}(0)=1$, and $\dot{b}_{\pm}=0$. The exact solutions in this case, for $0<t<\tau$, turns out to be:
\begin{align}\label{Eq:bp_sol}
	b_{+}(t)&=\sqrt{\frac{\omega^2 \sin ^2\left(\omega't\right)}{\omega'{}^2}+\cos^2\left(\omega't\right)}\,,\\\label{Eq:bm_sol}
b_{-}(t)&=\sqrt{\frac{\omega^2\sin^2\left(\eta\omega't\right)}{\eta^2\omega'{}^2}+\cos^2\left(\eta\omega't\right)}\,,
\end{align}
where, $\eta^2\equiv \frac{\omega'{}^2+2k}{\omega'{}^2}$.
In summary, the full quantum dynamics of the system, when prepared in the ground state at $t=0$, is contained in \ref{Eq:bp_sol} and \ref{Eq:bm_sol}. 

\begin{figure}[t!]
    \includegraphics[width=.9\linewidth]{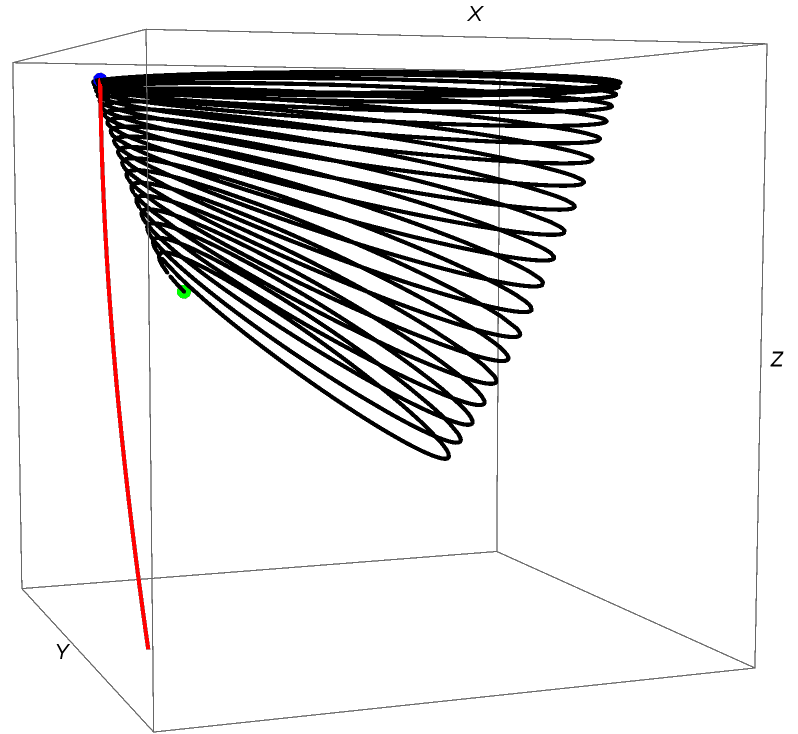}
    \caption{\justifying Evolution of $\rho^{(1)}_{x_1\,x'_1}$ represented in the $\vec{R}$-space, assuming a randomly chosen set of the tunable parameters $\omega'$, $k$ and $\tau$. The \textit{oscillator}-1 is initially in the ground state (blue dot) and evolves, at $t=\tau$, to the green point. The red curve is the family of thermal states. }
    \label{fig:randon_rho}
\end{figure}

To visualize the evolution of the density matrix that follows from the above solutions, we can derive the corresponding $\vec{R}(t)$ that follows from \ref{eq:def_R} and \ref{eq:def_A11}--\ref{Eq:bm_sol}. For arbitrary values of the tunable parameters $\omega'$, $k$, and $\tau$, the resulting evolution does not necessarily lead to the thermal state. To demonstrate this visually, in \ref{fig:randon_rho} we present the evolution of $\vec{R}(t)$ for a random choice of these parameters. Note again that the red curve represents the family of thermal states, and clearly, the density matrix does not evolve to this line in this specific example. In the next section, we shall show that by tuning the above parameters, we can prepare the reduced density matrix of one of the oscillators in a thermal form.

\begin{figure}[h]
 \centering
    \begin{subfigure}[t]{0.5\textwidth}
        \centering
    \makebox[0pt]{\includegraphics[width=.9\linewidth]{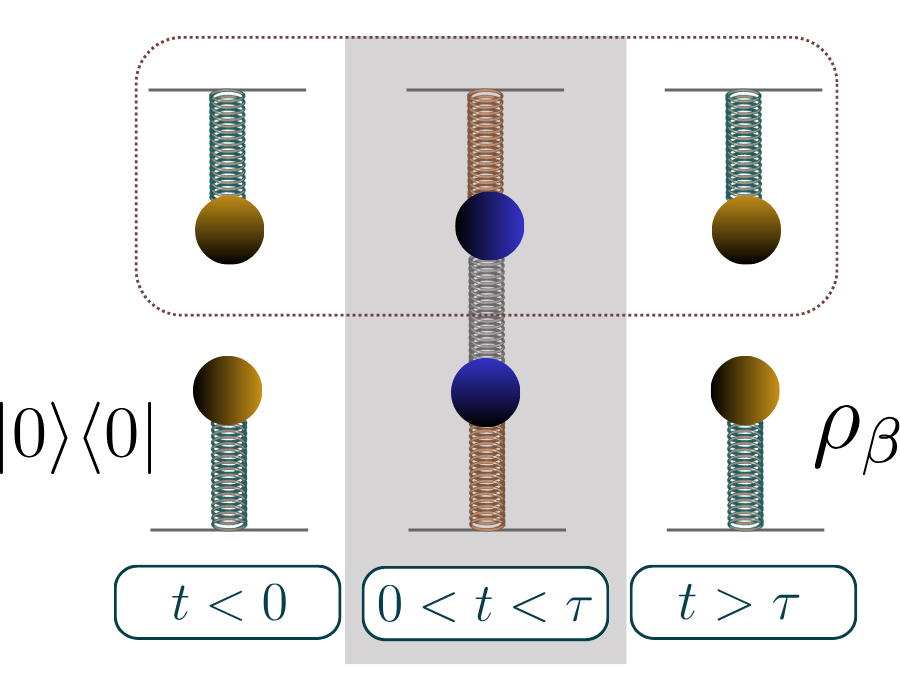}}
    \caption{}
    \end{subfigure}\\
    \begin{subfigure}[t]{0.5\textwidth}
        \centering
    \makebox[0pt]{\includegraphics[width=.9\linewidth]{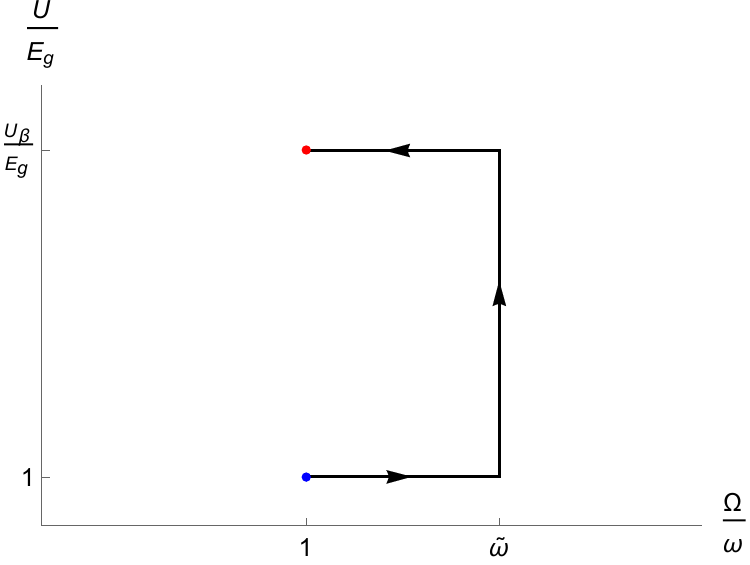}}
    \caption{}
    \label{fig:E_vs_omega}
    \end{subfigure}
    \caption{\justifying \textbf{(a)} Protocol for preparing \textit{oscillator}-1 in a thermal state. Both oscillators start in their ground states and are initially decoupled. A coupling of strength $k$ is suddenly introduced, and their frequencies are shifted to $\omega'$ for a duration $\tau$ (the \emph{active-phase}). The oscillators are then decoupled and their frequencies reset to $\omega$. Tuning $k$, $\omega'$, and $\tau$ brings \textit{oscillator}-1 to a thermal state at $t = \tau$. \textbf{(b)} The energy-frequency diagram for \textit{oscillator}-1 as it undergoes the protocol.}
    \label{fig:scheme}
\end{figure}

\section{Thermalization from quenching}\label{sec:main}

We begin with a brief schematic overview, as illustrated in \ref{fig:scheme}, demonstrating how our framework enables the preparation of \textit{oscillator}-1 in its thermal state. Recall that \textit{oscillator}-1 is the system of interest and is initially prepared in its ground state. The method also involves a second, identical oscillator (\textit{oscillator}-2), which is likewise initialized in the ground state and decoupled from \textit{oscillator}-1 initially. In the trapped-ion implementation of this setup, such ground-state preparation can be achieved, for example, via laser cooling, as in \cite{monroe1995resolved,meekhof1996generation}.

To rephrase the time dependence of the system as described by \ref{eq:defin_quench} in more physical terms, define an \emph{active-phase} during which the coupling of strength $k$ is suddenly switched on between the two oscillators, and their frequencies are abruptly shifted to a new value $\omega'$. This phase lasts for a duration $\tau$, after which the oscillators are decoupled and their frequencies are restored to the original value $\omega$, marking the end of the \emph{active-phase}. We will now show that, by suitably tuning $k$, $\omega'$, and $\tau$, \textit{oscillator}-1 evolves into a thermal state at time $t = \tau$. To this end, it is convenient to introduce the following dimensionless parameters:
\begin{align}
    \tilde{\omega}' \equiv \frac{\omega'}{\omega}, \quad 
    \tilde{k} \equiv \frac{k}{\omega^2}, \quad 
    \tilde{\tau} \equiv \frac{\omega\tau}{2\pi} \,,
\end{align}
which allows us to express the tunable quantities in terms of $\omega^{-1}$, which is the natural time scale of the oscillators in the uncoupled state.

In the previous section, we outlined how the reduced density matrix $\rho^{(1)}_{x_1\,x_1'}$ of \textit{oscillator}-1 evolves in our setup, showing that its dynamics can be effectively captured by the vector $\vec{R}(t)$. Therefore, the condition that \textit{oscillator}-1 thermalizes at $t=\tau$ can be written concisely as
\begin{align}\label{eq:thermality}
    \vec{R}(\tau)=\vec{R}_{\beta}\,.
\end{align}
In term of the solutions $b_{\pm}(t)$ of the Ermakov equations \ref{eq:ermakov}, the above condition gives rise a system of three algebraic equations for the unknowns $\tilde{k}$, $\tilde{\omega}'$, and $\tilde{\tau}$. The goal, then, is to solve the system for a given temperature $T=1/\beta$, in order to determine the appropriate values of $\tilde{k}$, $\tilde{\omega}'$, and $\tilde{\tau}$ that correspond to that temperature.

Although this system of equations does not appear to yield an exact analytical solution for arbitrary temperatures, it is, in principle, amenable to numerical solution, thereby confirming the theoretical feasibility of our protocol. Interestingly, however, we will now examine a \textit{special discrete set} of temperature values — dense in the positive real line — for which the otherwise complicated system does admit exact analytical solutions.

\subsection{Special discrete set of thermal states}

The classical dynamics of coupled oscillators display quasi-periodicity due to the presence of two normal modes frequencies, which are generally incommensurate. This characteristic persists in the quantum dynamics as well, as reflected in our setup by the forms of the solutions $b_{\pm}(t)$ in \ref{Eq:bp_sol} and \ref{Eq:bm_sol}. However, in the special case where the normal modes are commensurate, the system becomes periodic. By inspection, we found that in this periodic limit the system of equations \ref{eq:thermality}, which determines the value of tunable parameters for achieving thermalization, admits an exact analytical solution for a \textit{special discrete set}(SDS) of temperatures, as we shall now demonstrate. 

The condition of periodicity, during the \textit{active-phase}, means that $\tilde{\omega}'/\sqrt{\tilde{\omega}'{}^2+2\tilde{k}}$ is a rational number. Of special interest to us is the case where this ratio takes the form of an odd-over-odd fraction, i.e., $p/q$ where $p$ and $q$ are odd integers. Under these assumptions, we propose the choice
\begin{align}\label{eq:tune_tau}
    \tilde{\tau}=\frac{1}{4\tilde{\omega}'}(2l+1)=\frac{1}{4\sqrt{\tilde{\omega}'{}^2+2\tilde{k}}}(2n+1)\,,
\end{align}
where $l,n\in \mathbb{W}\,$. Physically, the first condition requires tuning $\tau$ to an odd multiple of one-quarter of the period associated with the normal mode $x_{+}$. The second line then \textit{follows} under our assumption that $\tilde{\omega}'/\sqrt{\tilde{\omega}'{}^2+2\tilde{k}}$ is an odd-over-odd fraction. The motivation behind our assumptions till \ref{eq:tune_tau} is that one of the conditions in \ref{eq:thermality}---namely, $Y(\tau)=0$--- is then automatically satisfied. 

To solve the rest of the equations, let us denote by $\eta$ the ratio of the normal mode frequencies in the \textit{active phase}:
\begin{align}
    \eta=\frac{\sqrt{\tilde{\omega}'{}^2+2\tilde{k}}}{\tilde{\omega}'}=\frac{2n+1}{2l+1}\,
\end{align}
where the second line follows from \ref{eq:tune_tau}. The thermalization condition, under the aforementioned assumptions, then reduces to
\begin{align}
    \frac{\left(\eta ^4+6 \eta ^2+1\right)}{(\eta^2-1)^2}&=\cosh(\omega\beta)\\
    \frac{4\left(\eta ^2+1\right)}{ \tilde{\omega}'{}^2\left(\eta ^2-1\right)^2}&=\sinh(\omega\beta)
\end{align}
It is easy to show that the above equations are solved by:
\begin{align}\label{Eq:beta_quantum}
	\frac{1}{ \tilde{\omega}'{}^2}=\eta= \left[\tanh\left(\frac{\beta\omega}{4}\right)\right]^{\pm 1}=\frac{2n+1}{2l+1}\,.
\end{align}
Further solving for $\tilde{\omega}'$ and $\tilde{k}$ explicitly, we arrive at the final result:
\begin{empheq}[box=\fbox]{align}\label{eq:tune_T}
  \frac{E_{g}}{k_{B}T_{nl}}&\equiv\begin{cases}
      2\tanh^{-1}\left(\frac{2l+1}{2n+1}\right)\,&;\,l<n\\
      2\coth^{-1}\left(\frac{2l+1}{2n+1}\right)\,&;\,l>n
  \end{cases}\,,\\\nonumber
  &=\log\left(\frac{n+l+1}{|n-l|}\right)\\\label{eq:tune_omegap}
  \tilde{\omega}'&= \sqrt{\frac{2l+1}{2n+1}}\equiv \tilde{\omega}_{nl}'\,,\\\label{eq:tune_k}
  \tilde{k}&=\frac{2(n-l)(n+l+1)}{(2l+1)(2n+1)}\equiv \tilde{k}_{nl}\,,\\\label{eq:tune_tau_final}
  \tilde{\tau}&=\frac{1}{4}\sqrt{(2l+1)(2n+1)}\equiv \tilde{\tau}_{nl} \,,
\end{empheq}
where $T_{nl}$ denotes the temperature and the last line is obtained by substituting the solutions of $(\tilde{\omega}',\tilde{k})$ into \ref{eq:tune_tau}. Note that we have retained $\hbar$ and $k_{B}$ momentarily so that \ref{eq:tune_T} is phrased in terms of the ground-state energy of \textit{oscillator}-1 ($E_{g}=\hbar\omega/2$) and the thermal energy scale ($k_{B}T_{nl}$). 

While a full discussion of the experimental details is reserved for future work, we provide rough estimates to give a sense of the relevant parameter space by considering two representative regimes at typical~\cite{Jefferts1995} MHz-scale trap frequencies: 1 MHz at 10 $\mu$K and 2 MHz at 25 $\mu$K. The numerical estimates are obtained by identifying suitable integer pairs $(l, n)$ satisfying the equations \ref{eq:tune_T} - \ref{eq:tune_tau_final}. For the 1 MHz regime, representative pairs $(288, 240)$ and $(294, 245)$ yield a quenched frequency $\omega' = 2\pi\times 1.10$ MHz, inter-mode coupling $k = -7.22 \times 10^{12}\,\mathrm{rad}^2\,\mathrm{s}^{-2}$, and interaction times $\tau = 131.7\,\mu\mathrm{s}$ and $\tau = 134.4\,\mu\mathrm{s}$ respectively, with error $\sim 10^{-4}\,\%$ in both cases. For the 2 MHz regime, pairs $(238, 177)$ and $(281, 209)$ give a quenched frequency $\omega' = 2\pi\times2.32$ MHz, coupling $k = -4.73 \times 10^{13}\, \mathrm{rad}^2\, \mathrm{s}^{-2}$, and interaction times $\tau = 51.4\,\mu\mathrm{s}$ and $\tau = 60.7\,\mu\mathrm{s}$ respectively, with  error $\sim 10^{-4} \,\%$ in both cases. Pairs with lower indices and smaller differences, such as $(11, 12)$ or $(22, 24)$, are expected to be easier to realize experimentally as they demand more modest RF modulation and shorter coherence times. This also implies that lower target temperatures, which correspond to these closely spaced pairs, may be more readily achieved.

\begin{figure}[t]
    \centering
    \begin{subfigure}[t]{0.5\textwidth}
        \centering
        \includegraphics[width=.9\textwidth]{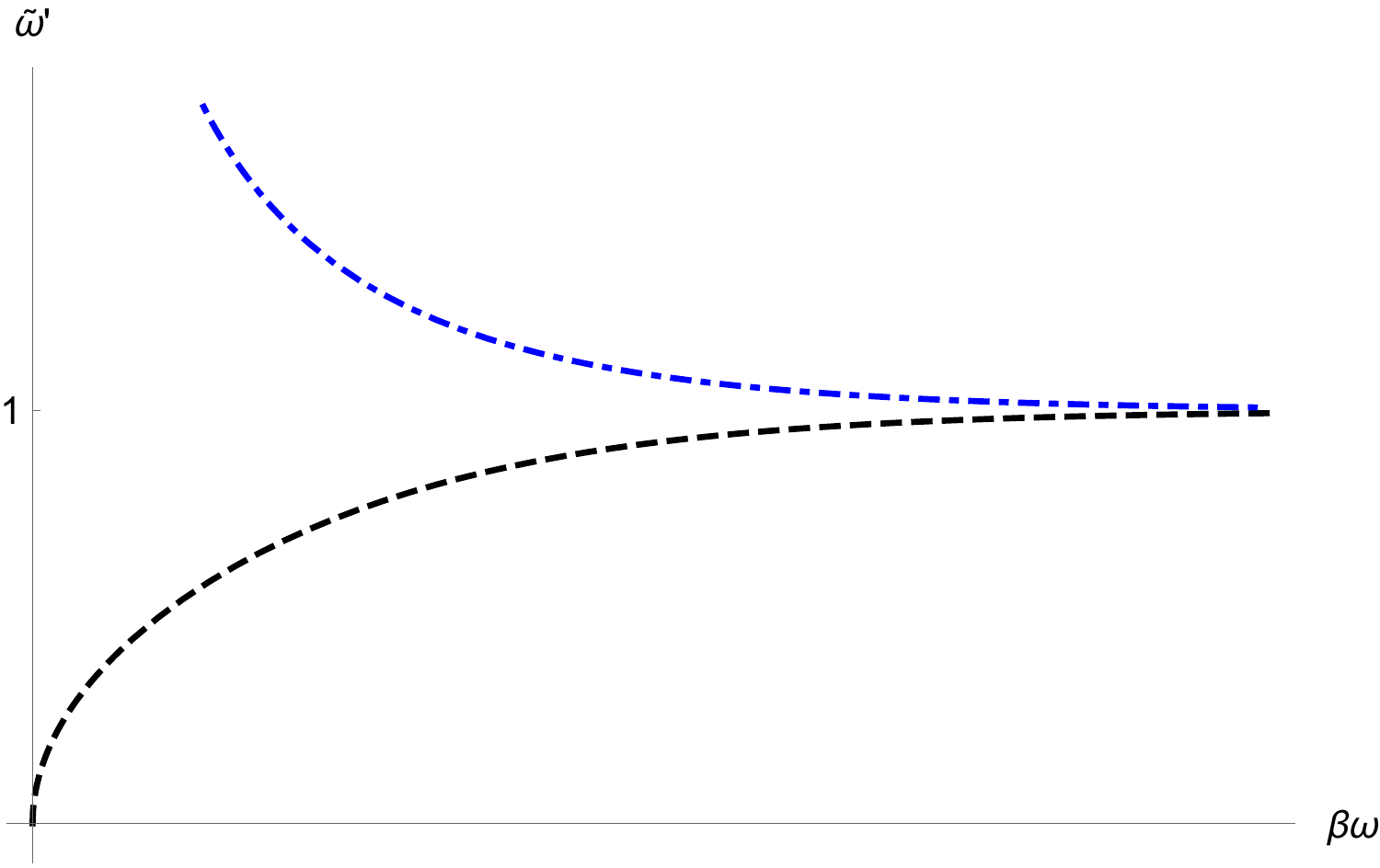}
       \caption{}
    \end{subfigure}%
    \\
    \begin{subfigure}[t]{0.5\textwidth}
        \centering
        \includegraphics[width=.9\textwidth]{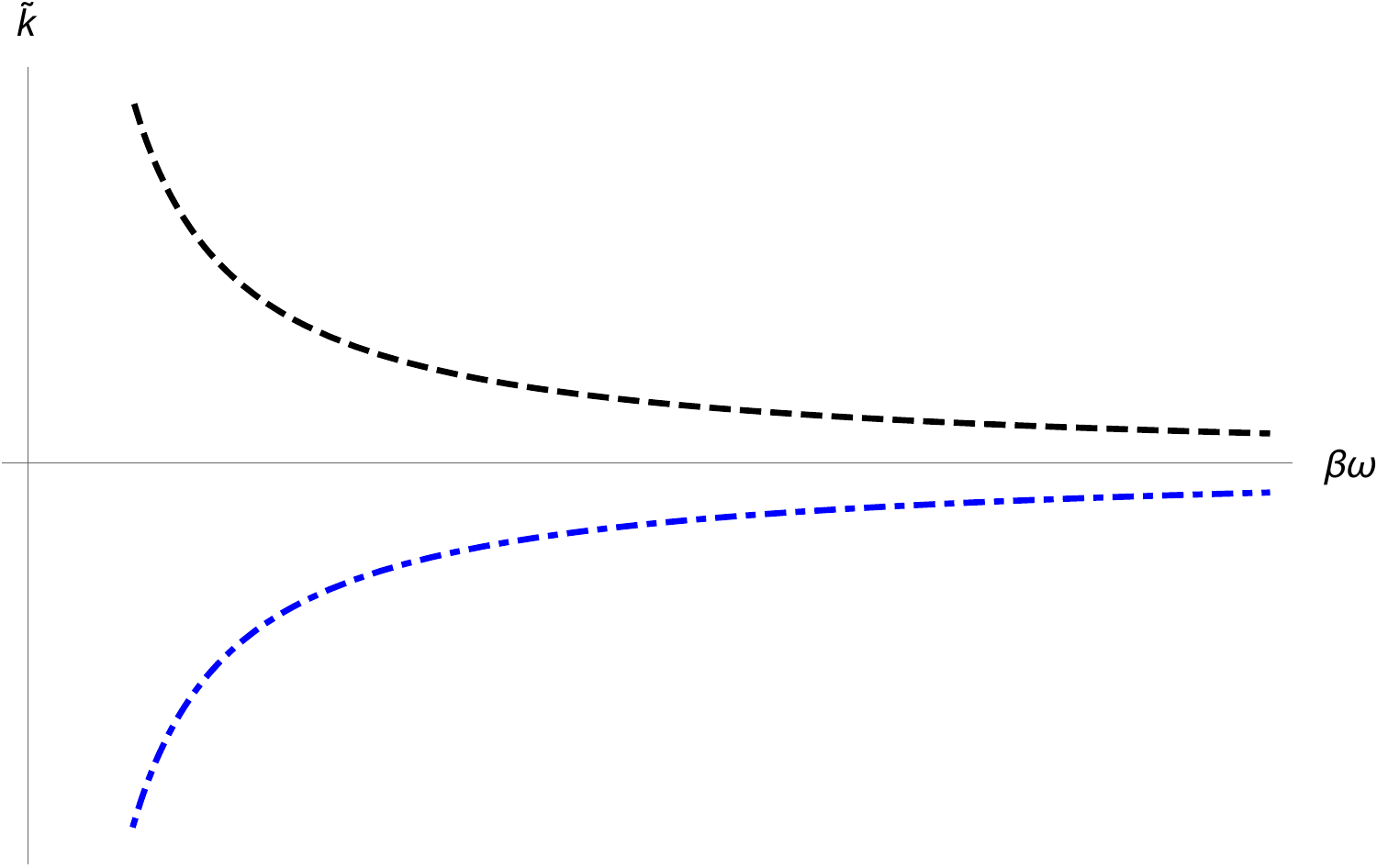}
        \caption{}
    \end{subfigure}
    \caption{\justifying The tunable parameters $\tilde{\omega}'$ and $\tilde{k}$ as a function of the SDS (inverse-)temperatures and as given in \ref{eq:omegapr_vs_T} and \ref{eq:k_vs_T}. The dashed and dot-dashed curves correspond to $l<n$ and $l>n$ cases, respectively. Note that these relations are valid only on the SDS $T_{nl}$. }\label{figure:tune_vs_T}
\end{figure}

Now, let's take a moment to gain a clearer understanding of the result. The first condition describes the SDS of temperatures identified by us as allowing for exact solutions of the thermalization condition \ref{eq:thermality}, as stated in the title of this subsection. This set is parametrized by two non-negative integers $l$ and $n$. In term of the the average thermal energy $U_{\beta}$ we can describe this set by rewriting \ref{eq:tune_T} in the form
\begin{align}
    U_{\beta_{nl}}=E_{g}\left[\frac{(2l+1)^2+(2n+1)^2}{2(2l+1)(2n+1)}\right]\,,
\end{align}
which is also manifestly symmetric under  $l\leftrightarrow n$. Therefore, there is a sort \textit{degeneracy} in that interchanging $l\leftrightarrow n$ maps to the same temperature.  As an example, consider $\beta=\frac{k_{B}}{E_g}\log\,2$, for which we can identify $(l,n)$ with either $(0,1)$ or $(1,0)$. Consequently, there are two inequivatent sets of tunable parameters $(\tilde{\omega}',\tilde{k},\tilde{\tau})$, namely, $(1/\sqrt{3},4/3,\sqrt{3}/4)$ and $(\sqrt{3},-4/3,\sqrt{3}/4)$, that give rise to the same inverse-temperature $\beta=\frac{k_{B}}{E_g}\log\,2$. In general, under $l\leftrightarrow n$, the tunable parameters transform as $\tilde{\omega}'\rightarrow 1/\tilde{\omega}'$, $\tilde{k}\rightarrow -\tilde{k}$ and $\tilde{\tau}\rightarrow \tilde{\tau}$, as can easily be verified using \ref{eq:tune_omegap}, \ref{eq:tune_k} and \ref{eq:tune_tau_final}. In fact, at the level of the evolution of state of the full system, the transformation $l\leftrightarrow n$, effects $b_{+}(t)\leftrightarrow b_{-}(t)$, under which the reduced density matrix remains invariant.  Another useful way of expressing the degeneracy is to give the expressions of $\tilde{\omega}'$ and $\tilde{k}$ as a function of the temperature.
\begin{align}\label{eq:omegapr_vs_T}
    \tilde{\omega}'&=\begin{cases}
        \sqrt{\tanh\left(\frac{E_{g}}{2k_{B}T_{nl}}\right)}\quad;\quad l<n\\
        \sqrt{\coth\left(\frac{E_{g}}{2k_{B}T_{nl}}\right)}\quad;\quad l>n
    \end{cases}\\\label{eq:k_vs_T}
    \tilde{k}&=\begin{cases}
        \textrm{cosech}\left(\frac{E_{g}}{k_{B}T_{nl}}\right)\quad;\quad l<n\\
        -\textrm{cosech}\left(\frac{E_{g}}{k_{B}T_{nl}}\right)\quad;\quad l>n
    \end{cases}
\end{align}
The envelope curves illustrating these relations are shown in \ref{figure:tune_vs_T}. The plots are consistent with the intuitive expectation that achieving higher temperatures necessitates a more dramatic change in the oscillator frequency and a stronger coupling during the \textit{active phase}.

We emphasize again that our protocol enables \textit{exact} thermalization in a finite time. To quote the quickest case allowed by the SDS solutions, consider again $\beta=\frac{k_{B}}{E_g}\log\,2$, where we can choose $(l,n)$ to be either $(1,0)$ or $(0,1)$. The time taken to thermalize, in dimensionless units, is then given by $\tilde{\tau}=\sqrt{3}/4\approx .433$. In physical terms, this means that exact thermalization from the ground state occurs in about $43.3\%$ of the oscillator's natural time-period. 

 The SDS temperatures, as such, do not cover all the possible values of temperature. However, it is worth emphasizing that this set is a countably dense subset of the set of all temperatures, which corresponds to the positive real line. This follows in view of \ref{eq:tune_T} and the fact that odd-over-odd rational numbers form a countably dense subset of the real line. What this means is that for any given temperature $T$ and an arbitrarily small error bar $\delta T$, one can always find a pair $(l,n)$ such that
\begin{align}
    T-\delta T<T_{ln}<T+\delta T\,.
\end{align}
Since practically there is always an error bar or tolerance in an experimental context, by utilizing the SDS solutions, in principle, our protocol can, in fact, be used to \textit{approximate} thermalization to \textit{arbitrary} values of temperature.  

An important caveat in approximating a target temperature using the nearest value from the SDS is that achieving higher accuracy typically requires large $l$ and $n$, which, from Eq. \ref{eq:tune_tau_final}, entails longer thermalization times. This can be problematic when experimental constraints or resource limits impose an upper bound on $\tau$. Moreover, increasing the interaction time also enhances the system’s exposure to environmental noise; this is further discussed in \ref{subsec:noise}. 

However, faster thermalization may be achievable if we move beyond the SDS restriction. As a preliminary step in this direction, let $\vec{R}(\tilde{\omega},\tilde{k},\tilde{\tau})$ denote the components of the $\vec{R}$-vector for parameters $\vec{P}=(P^{1},P^{2},P^{3})\equiv(\tilde{\omega},\tilde{k},\tilde{\tau})$. For a desired inverse temperature $\beta=\beta_{nl}+\delta\beta$ near an SDS value, we can obtain an approximate perturbative correction to the control parameters by solving
\begin{align}
\sum_{j=1}^{3}\mathcal{M}^{i}_{\,\, j}\delta P^{j}
= \delta\beta\,\frac{\partial R^{i}_{\beta_{nl}}}{\partial \beta},
\end{align}
where
\begin{align}
\mathcal{M}^{i}_{\,\, j} \equiv \frac{\partial R^{i}(\vec{P}_{nl})}{\partial P^{j}},
\end{align}
and $R^{i}_{\beta}$ are given in \ref{eq:R_beta}. For $\det(\mathcal{M})\neq 0$, the first-order solution, with $P^{i}=P_{nl}^{i}+\delta P^{i}$, reads
\begin{align}
\delta P^{i}=
 \delta\beta\left(\mathcal{M}^{-1}\right)^{i}_{\,\, j}
\frac{\partial R^{j}_{\beta_{nl}}}{\partial \beta}
+ \mathcal{O}(\delta\beta^{2}).
\end{align}
Studying these linear perturbative solutions could shed light on the \textit{local} structure of the thermalization condition’s solution space. A complete characterization, however, requires a full numerical analysis of \ref{eq:thermality}, which we leave for future work.

We now turn to analyzing the system's evolution toward the thermal state within our protocol.

\subsubsection{Evolution to the thermal state}

\begin{figure}[h]
    \includegraphics[width=0.75\linewidth]{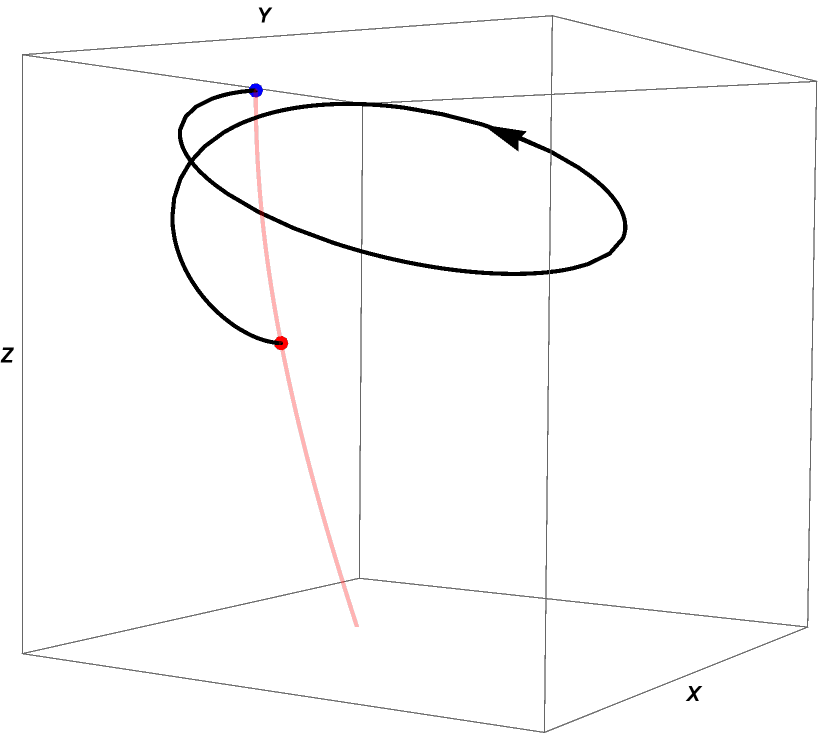}
    \caption{
        \justifying Finite-time thermalization of oscillator 1 to the thermal state at $\beta=\frac{k_{B}}{E_g}\log\,2$, starting from the ground state.}
    \label{fig:Rplot_quickest}
\end{figure}

To visualize the time evolution of the density matrix as the system thermalizes under our protocol, we once again invoke the $\vec{R}$-space defined in \ref{sec:gaussian}. As a first example, we revisit $\beta=\frac{k_{B}}{E_g}\log\,2$, corresponding to the quickest case in the SDS. The evolution of \textit{oscillator}-1 from the ground state to the thermal state at this temperature is plotted in \ref{fig:Rplot_quickest}.

\begin{figure*}[h!]
    \centering
    \begin{subfigure}[t]{0.475\textwidth}
        \centering
        \includegraphics[width=.9\textwidth]{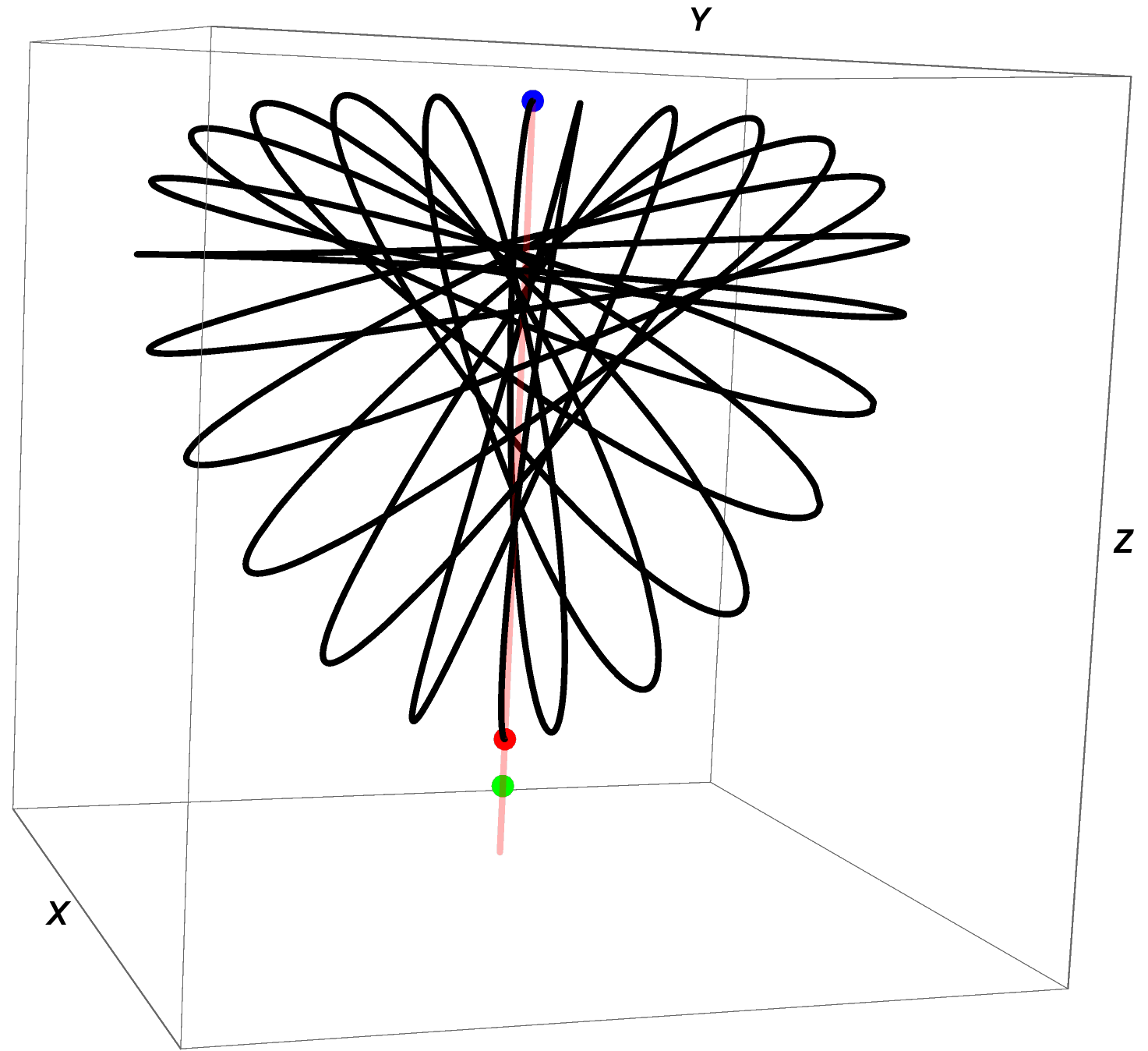}
       \caption{}
    \end{subfigure}%
    \begin{subfigure}[t]{0.475\textwidth}
        \centering
        \includegraphics[width=.9\textwidth]{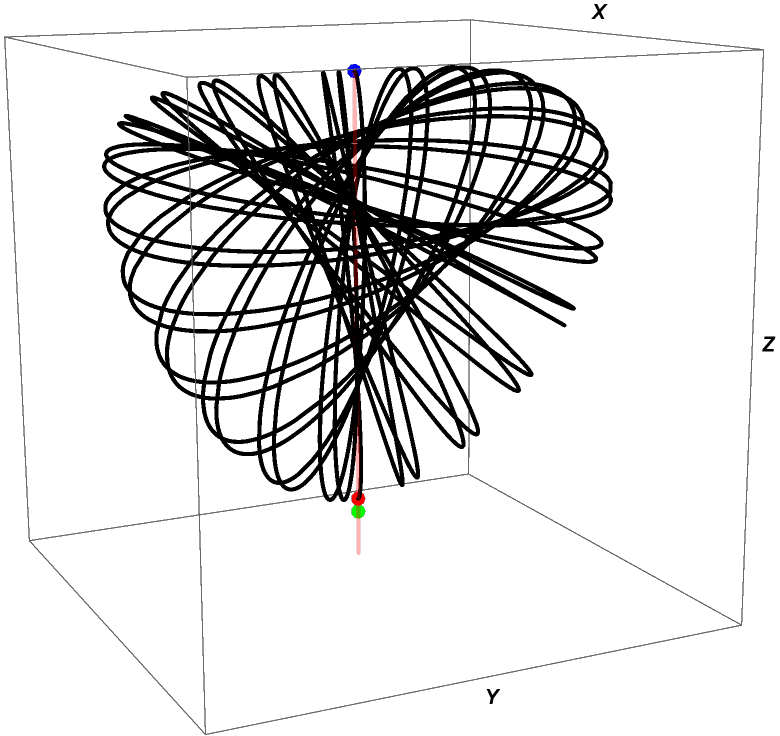}
        \caption{}
    \end{subfigure}  
\vskip\baselineskip
    \begin{subfigure}[t]{0.475\textwidth}
        \centering
        \includegraphics[width=.9\textwidth]{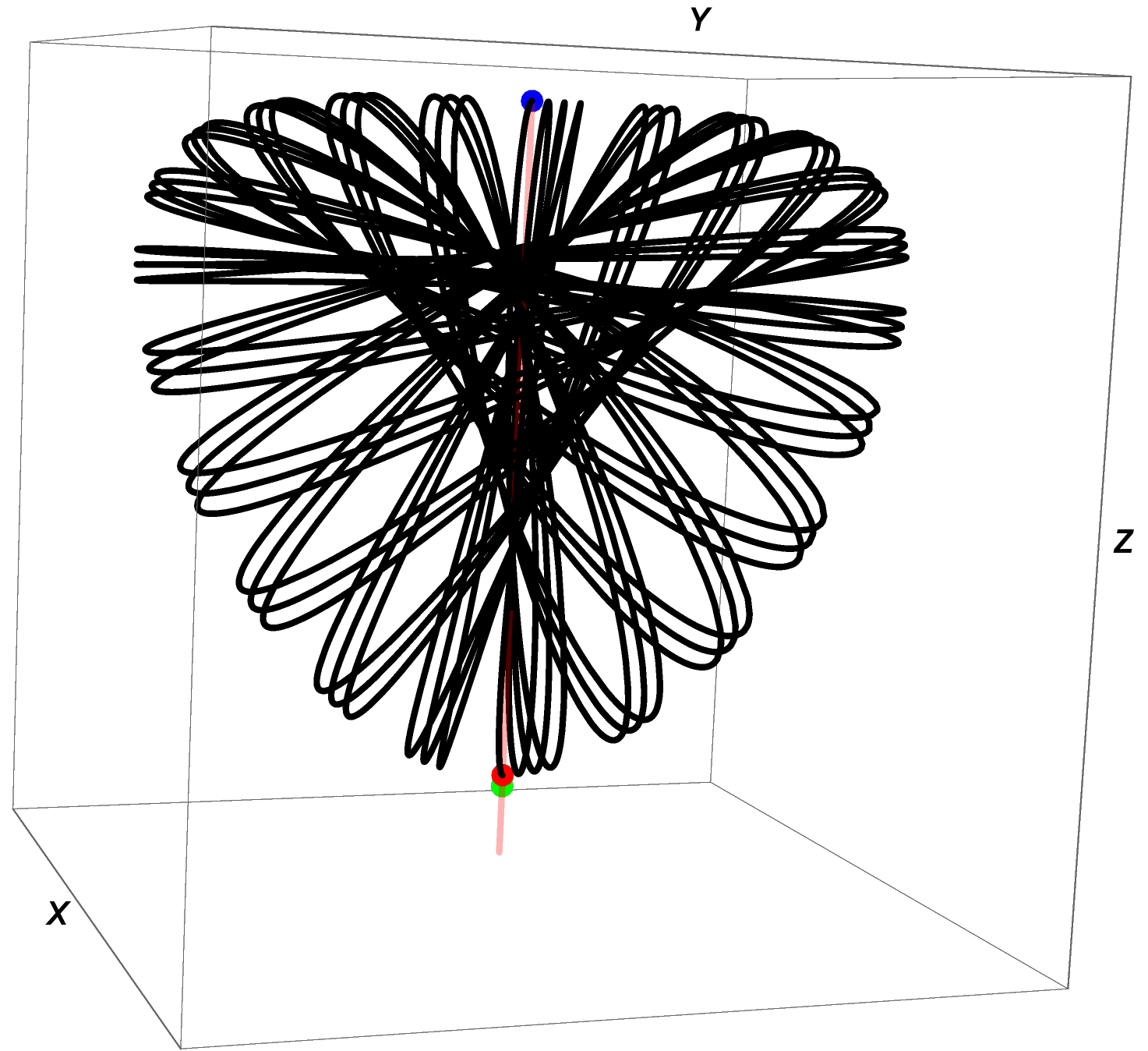}
        \caption{}
    \end{subfigure}
    \begin{subfigure}[t]{0.475\textwidth}
        \centering
        \includegraphics[width=.9\textwidth]{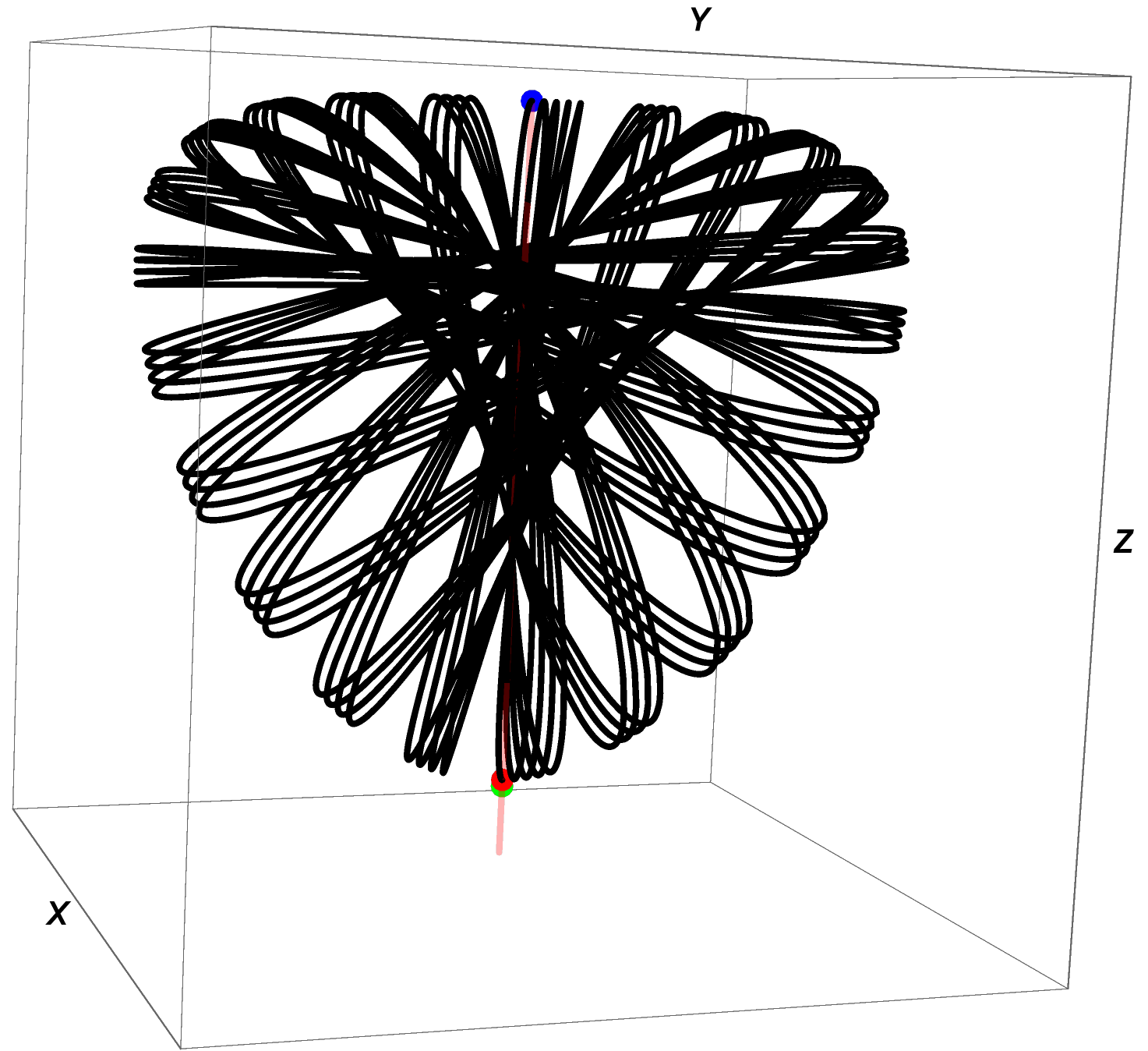}
        \caption{}
    \end{subfigure}
    \caption{\justifying Evolution toward the thermal state for various approximations to the target inverse temperature $\beta=\frac{k_{B}}{E_g}\pi$. The exact thermal state corresponding to this value is marked by a green dot, while the approximate thermal state is indicated by a red dot; both lie on the thermal curve.  The associated $(l,n)$ pairs are  $(12,11)$, $(24,22)$, $(36,33)$, and  $(48,44)$ for panels \textbf{(a)}, \textbf{(b)}, \textbf{(c)}, and \textbf{(d)}, respectively.}\label{figure:R_plot_pi}
\end{figure*}

Next, to illustrate a case where the target temperature \textit{does not} belong to the SDS, consider $\beta=\frac{k_{B}}{E_g}\pi$. To a first approximation--accurate within about $1.16\%$--this temperature can be matched using the pair $(l=11,n=12)$ or, equivalently, $(l=12,n=11)$, both yielding $\beta\approx\frac{k_{B}}{E_g}\times3.178$. However, as we remarked earlier, one can always choose $(n,l)$ values to match the desired temperature value of $\beta=\frac{k_{B}}{E_g}\pi$ with arbitrary accuracy. To illustrate this, we have listed a few such choices in \ref{table:nl_for_pi}, with the accuracy increasing as we move down the table. The corresponding evolution to the thermal state is presented in \ref{figure:R_plot_pi}. As we pointed out earlier, the increased accuracy comes at the cost of increased time duration required for thermalization. 

\begin{table}[h!]
\begin{center}
\begin{tabular}{ ||w{c}{1.5cm}|w{c}{1.5cm}|w{c}{1.5cm}|w{c}{1.5cm}||  }
 \hline
 $(l,n)$ & $E_g\beta_{nl}/k_{B}$& $\delta\beta/\beta$ & $\tilde{\tau}$ \\ [.2ex] 
 \hline\hline
 (12,11) & 3.178 & 1.16\% & 5.99 \\ 
 (24,22) & 3.157 & .49\% & 11.74 \\
 (36,33) & 3.149 & .26\% & 17.48 \\
 (48,44) & 3.146 & .15\% & 23.23 \\
 (60,55) & 3.144 & .08\% & 28.97 \\ [1ex] 
 \hline
\end{tabular}
\caption{\justifying List of SDS approximations to $\beta=\pi k_{B}/E_{g}$. As one moves down the list, the approximation error in temperature decreases by roughly a factor of two with each step. However, this improved accuracy comes at the cost of longer thermalization times, which in this example is scaling roughly inversely with the \%-error in temperature.}
\label{table:nl_for_pi}
\end{center}
\end{table}

\begin{figure}
 \centering
    \begin{subfigure}[t]{0.475\textwidth}
    \includegraphics[width=0.75\linewidth]{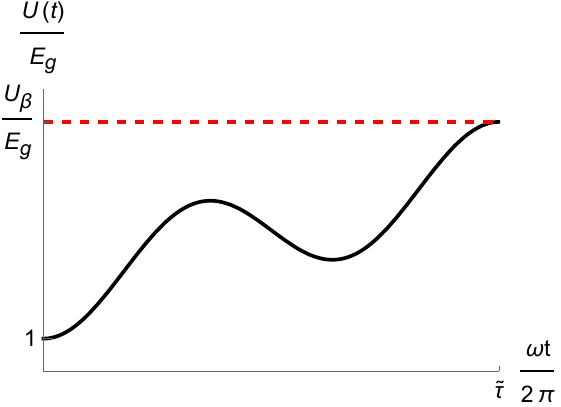}
    \caption{}
    \end{subfigure}\\
    \centering
    \begin{subfigure}[t]{0.475\textwidth}
     \includegraphics[width=.75\linewidth]{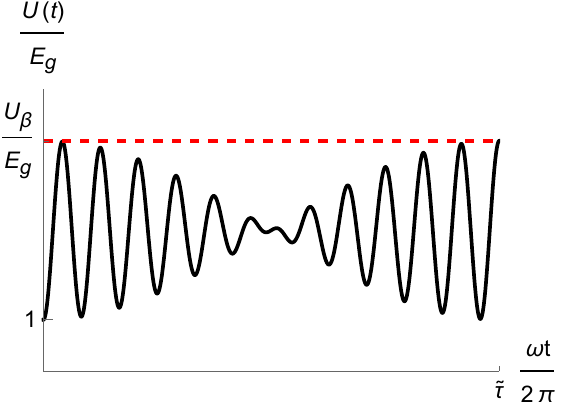}      
     \caption{}
     \end{subfigure}
    \caption{
\justifying
Time evolution of $U(t)$ shown as a solid black curve. The final thermal value is indicated by a dashed red line. \textbf{(a)} $\beta = \frac{k_B}{E_g} \log 2$, corresponding to $(l,n) = (1,0)$, and \textbf{(b)} $\beta\approx\frac{k_{B}}{E_g}\times3.178$, corresponding to $(l,n) = (12,11)$. The thermal energy $U_\beta$ for $\beta = \frac{k_B}{E_g} \pi$ is also displayed in light red.
}
 \label{fig:U_vs_t}
\end{figure}

Another way to study the thermalization process is by analyzing the time evolution of the average energy of the system—specifically, \textit{oscillator}-1—as it approaches thermal equilibrium. The average energy \( U(t) \) is defined as
\begin{align}
    U(t) &= \braket{H_0} = \mathrm{Tr}\left[ \rho^{(1)} H_0 \right], \\
    H_0 &= \frac{p_1^2}{2m} + \frac{1}{2} m \omega^2 x_1^2\,,
\end{align}
where \( H_0 \) is the Hamiltonian of the first oscillator. A direct evaluation of the expectation value yields
\begin{align}
    U(t) = \frac{E_{g}}{2}\frac{\left(1 + X^2 + Y^2-Z^2 \right)}{(X+Z)}\,,
\end{align}
where the expression is written in terms of the components of \( \vec{R} \). The evolution of the average energy toward the thermal value \( U_\beta \) is shown in \ref{fig:U_vs_t} for two of the cases previously discussed: \( (l,n) = (1,0) \) and \( (12,11) \). The observed oscillatory behavior is expected, given the structure of \( \rho^{(1)} \) in our setup.

A related quantity of interest is the time evolution of the von Neumann entropy $\mathcal{S}(t)$ of  \textit{oscillator}-1 as it thermalizes. The mixed nature of its reduced density matrix arises from entanglement with the second oscillator. In particular, this means that the thermal entropy of \textit{oscillator}-1 is simply the entanglement entropy at the end of the protocol.

\begin{figure}
 \centering
    \begin{subfigure}[t]{0.475\textwidth}
    \includegraphics[width=0.75\linewidth]{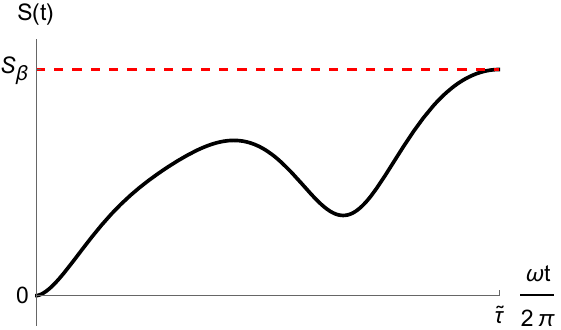}
    \caption{}
    \end{subfigure}\\
    \centering
    \begin{subfigure}[t]{0.475\textwidth}
     \includegraphics[width=.75\linewidth]{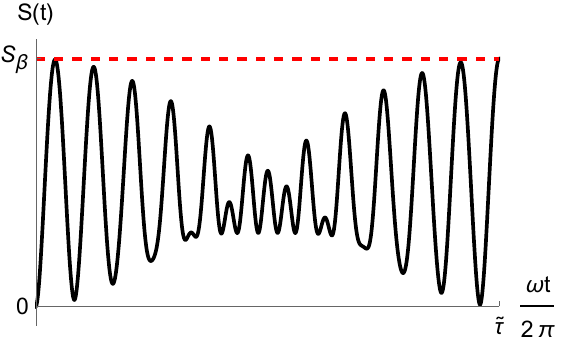}      
     \caption{}
     \end{subfigure}
    \caption{
\justifying
{The time evolution of $S(t)$ is shown as a solid black curve, with the corresponding final thermal value indicated by a dashed red line. Panel \textbf{(a)} corresponds to $(l,n) = (1,0)$, with thermal entropy $S_{\beta}$ at $\beta = 2 \log 2$. Panel \textbf{(b)} corresponds to $(l,n) = (12,11)$, with thermal entropy $S_{\beta}$ at $\beta = 2 \log 24$, shown in red.
}}
 \label{fig:S_vs_t}
\end{figure}

The von Neumann entropy for $\rho^{(1)}$ can be expressed in terms of the simplectic eigenvalue $\nu$ of the covariant matrix $\Sigma$ as
\begin{align}
\mathcal{S}=\left(\nu+\frac{1}{2}\right)&\log\left(\nu+\frac{1}{2}\right)\\\nonumber
&-\left(\nu-\frac{1}{2}\right)\log\left(\nu-\frac{1}{2}\right) \,. 
\end{align}
The symplectic eigenvalue is given by
\begin{align}
    \nu=\sqrt{\det(\Sigma)}=\frac{1}{2}\sqrt{\frac{X^2+Y^2-Z^2}{(X+Z)^2}}\,,
\end{align}
which is related to purity $\mu$ simply by $\nu=1/(2\mu)$. For comparison, the thermal entropy $\mathcal{S}_{\beta}$ takes the form
\begin{align}
    \mathcal{S}_{\beta}=\frac{\beta\omega}{e^{\beta\omega}-1}-\log\left(1-e^{-\beta\omega}\right)\,.
\end{align}
Paralleling our preceding discussion of the average energy $U(t)$, the evolution of the von Neuman entropy toward  \( \mathcal{S}_\beta \) is shown in \ref{fig:S_vs_t} for the two cases \( (l,n) = (1,0) \) and \( (12,11) \).

\subsubsection{Environmental effects}\label{subsec:noise}

We have seen that achieving a more accurate SDS approximation of a given temperature typically requires longer protocol durations. A direct consequence of this increased duration is a greater sensitivity to environmental coupling effects accumulated during the finite interaction time. In the following, we quantify this sensitivity within a Gaussian Markovian open-system framework, where deviations from ideal unitary dynamics arise from a Lindblad-type coupling to an external bath.

The non-unitary evolution of a Gaussian system is fully characterized by the dynamics of its covariance matrix $\mathbf{\Sigma}$ (denoted in bold to distinguish it from the subsystem covariance matrix $\Sigma$ used in \ref{eq:def_sigma}). Following \cite{Mancini:2006pju}, a simple linear model for such dynamics takes the form
\begin{align}\label{eq:EOM_Sig}
    \dot{\mathbf{\Sigma}}=\A\mathbf{\Sigma}+\mathbf{\Sigma}\A^{T}+\mathbf{D}\,,
\end{align}
where $\A$ and $\mathbf{D}$ are the drift and diffusion matrices, respectively. These matrices can be written as
\begin{align}
    \A&=\A_{H}+\A_{\Gamma}, \qquad 
    \A_{\Gamma}=\J\,\mathrm{Im}[\Gamma],\\
    \mathbf{D}&=\J\,\mathrm{Re}[\Gamma]\,\J^{T}\,,
\end{align}
where $\J$ is the symplectic form, $\A_{H}$ generates the unitary Hamiltonian evolution, and $\Gamma=C^{\dagger}C$ with $C$ being a constant coupling matrix.

To quantify the effect of the non-unitary contribution, we consider the evolution of the log-purity, which is proportional to $\log\det(\Sig)$. We separate the unitary dynamics by defining $\Sig=e^{\A_{H}t}\tilde{\Sig}e^{\A_{H}^{T}t}$ to rewrite \ref{eq:EOM_Sig} as
\begin{align}\label{eq:EOM_Sig_2}
    \frac{d}{dt}\log\det(\Sig)&= 2\textrm{Tr}[\A_{\Gamma}(t)]+\textrm{Tr}[\tilde{\Sig}^{-1}\mathbf{D}(t)]\,,
\end{align}
where we define
\begin{align}
    \A_{\Gamma}(t)&\equiv e^{-\A_{H}t}\A_{\Gamma}e^{\A_{H}t}\,,\textrm{ and}\\
     \mathbf{D}(t)&\equiv e^{-\A_{H}t}\mathbf{D}e^{-\A_{H}^{T}t}\,.
\end{align}

The unitary evolution matrices $e^{-\A_{H}t}$ and $e^{-\A_{H}^{T}t}$ can be computed explicitly for the present two-oscillator system.  However, for our purposes it is sufficient to note that their matrix elements reduce to linear combinations of the functions $c_{\pm}(t)=\cos(\Omega_{\pm}t)$ and $s_{\pm}(t)=\sin(\Omega_{\pm}t)$, a structure that becomes most transparent in the normal-mode basis. 

To leading order in the system–bath coupling, we treat the interaction perturbatively and approximate $\tilde{\mathbf{\Sig}}(t)\simeq \mathbf{\Sigma}(0)$ in \ref{eq:EOM_Sig_2}. Assuming an initially pure Gaussian state, the change in log-purity $\propto\Delta \log\det \mathbf{\Sigma}(\tau)$, then separates into two qualitatively distinct contributions: (i) secular terms arising from time integrals of oscillatory functions such as $\int_{0}^{\tau} c_{\pm}^2(t)\,dt$, which grow at most linearly with $\tau$, and (ii) purely oscillatory terms involving bounded combinations of sines and cosines, which remain uniformly bounded in time. As a consequence, one obtains
\begin{align}
    \big|\Delta \log\det \mathbf{\Sigma}(\tau)\big| \sim \mathcal{C}_1 \tau + \mathcal{C}_2(\tau),
\end{align}
with $\mathcal{C}_{1}$ being a constant $\mathcal{C}_2(\tau)$ a bounded function, both determined by the system–bath coupling strength and the initial covariance matrix. In particular, $\mathcal{C}_1^{-1}$ defines the characteristic decoherence timescale. This establishes that the sensitivity to environmental coupling grows with the protocol duration in the stable Gaussian regime, providing a direct trade-off between thermalization accuracy and noise accumulation.

This concludes our discussion on employing the proposed setup to thermalize a harmonic oscillator initially prepared in its ground state. Interestingly, as a natural extension of the protocol, the same setup can also be used to either heat or cool \textit{oscillator}-1. Before concluding, we now briefly outline how this can be achieved.

\subsection{Corollary: heating and cooling from quenching }

We begin by recalling our earlier observation that the SDS arises when the normal mode frequencies of the \textit{active phase} are commensurate. This condition ensures that the system's evolution is periodic. A remarkable consequence of this is that the same quench protocol that we introduced in the previous section can also be employed to \emph{cool} \textit{oscillator}-1—initially prepared in a thermal state corresponding to a temperature within the SDS—down to its ground state. In the energy-frequency diagram, this scenario corresponds to a reversal of the arrows shown in \ref{fig:E_vs_omega}.

As an illustrative example, consider the total system initially prepared in the purification of \textit{oscillator}-1's thermal state at inverse temperature \(\beta = \frac{k_B}{E_g} \log 2\). Applying our protocol with parameters \((\tilde{\omega}', \tilde{k}, \tilde{\tau}) = \left(1/\sqrt{3}, 4/3, \sqrt{3}/4 \right)\) evolves the system into its ground state by the end of the \textit{active phase}.

\begin{figure}
 \centering
    \begin{subfigure}[t]{0.5\textwidth}
        \centering
    \makebox[0pt]{\includegraphics[width=.8\linewidth]{schematic_heat.pdf}}
    \caption{}
    \end{subfigure}\\
    \begin{subfigure}[t]{0.5\textwidth}
        \centering
    \makebox[0pt]{\includegraphics[width=.7\linewidth]{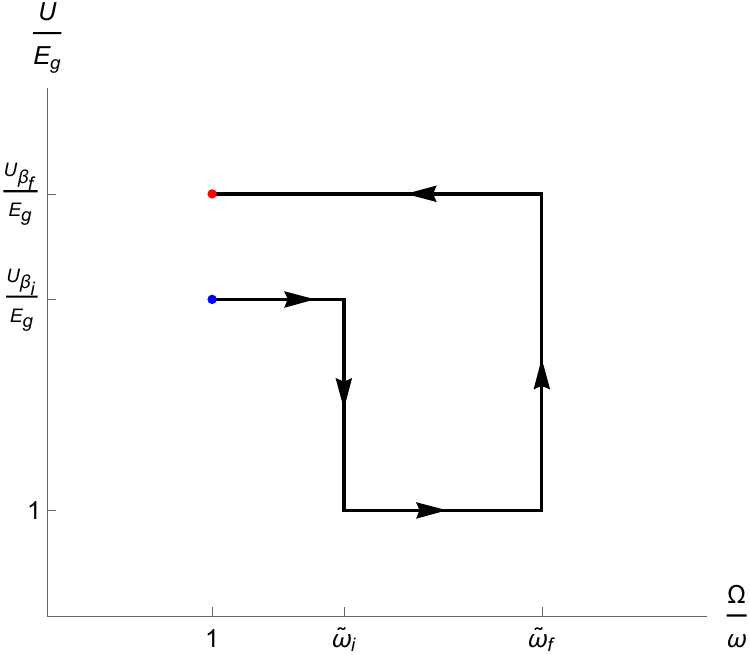}}
    \caption{}
    \label{fig:E_vs_omega_heating}
    \end{subfigure}
   \caption{\justifying  
\textbf{(a)} Adaptation of our protocol to enable either heating or cooling of \textit{oscillator}-1. \textbf{(b)} The corresponding energy-frequency diagram, shown here for the case of heating.
}

    \label{fig:scheme_heating}
\end{figure}

Taking this a step further, by applying two such quench protocols in sequence, one can effectively heat or cool the oscillator—that is, evolve it from an initial inverse temperature \(\beta_i\) to a final one \(\beta_f\). This amounts to the following time dependence of the frequency and coupling:
\begin{align}\label{eq:defin_quench_twice}
	\mathcal{K}(t)&=\begin{cases}
	\omega^2\,\tilde{k}_i\quad &;\quad 0<t<\tau_{i}\\
    \omega^2\,\tilde{k}_{f}\quad&;\quad\tau_{i}< t<\tau_i+\tau_{f}\\
	0\quad &;\quad \textrm{otherwise}
	\end{cases}\,,\\\nonumber
	\Omega(t)&=\begin{cases}
	\omega\,\tilde{\omega}'_i\quad &;\quad 0<t<\tau_{i}\\
    \omega\,\tilde{\omega}'_{f}\quad&;\quad\tau_{i}< t<\tau_i+\tau_{f}\\
	\omega\quad &;\quad \textrm{otherwise}
	\end{cases}\,,
\end{align}
where the parameter sets \((\tilde{\omega}'_{i}, \tilde{k}_{i}, \tilde{\tau}_{i})\) and \((\tilde{\omega}'_{f}, \tilde{k}_{f}, \tilde{\tau}_{f})\) should be chosen according to \ref{eq:tune_T}--\ref{eq:tune_tau_final}, so as to correspond to the inverse temperatures \(\beta_{i}\) and \(\beta_{f}\), respectively. A schematic of this process, along with the corresponding energy-frequency diagram, is shown in \ref{fig:scheme_heating}.

An interesting direction for future research is to examine the efficiency of the heating and cooling processes discussed above and to analyze their thermodynamic implications—for instance, by extending the single-oscillator treatment in \cite{deffner2008nonequilibrium} to our coupled system, or by exploring potential relevance to quantum thermal control \cite{Jussiau_2021}. While we have shown that the quench sequence can successfully achieve both heating and cooling, it is likely not the most efficient method within our setup. A promising avenue for future work is to optimize these processes in coupled oscillator systems through time-dependent control of the frequency and coupling parameters, employing techniques from optimal control theory \cite{kirk2004optimal}.

\section{Discussion}\label{sec:discussion}

In this work, we have proposed and analyzed a simple, finite-time protocol for preparing a quantum harmonic oscillator in a thermal state starting from its ground state. By replacing a conventional macroscopic heat bath with a second oscillator and using a short sequence of quenches in the oscillator frequencies and coupling strength, we have shown that exact thermalization can be engineered. The protocol relies on the Gaussian nature of the dynamics, which allows the reduced state of the target oscillator to be tracked compactly through the three-parameter vector $\vec{R}(t)$. 

In the present work, we demonstrated that the thermalization condition reduces to a set of three equations in the tunable parameters $(\tilde{\omega}',\tilde{k},\tilde{\tau})$. While these equations are in general solvable numerically for arbitrary temperatures, we identified a \emph{special discrete set} of temperatures for which the system admits exact, closed-form solutions. This set, labelled by two integers $(l,n)$, is countably dense in the positive real line of temperatures, implying that any target temperature can be approximated to arbitrary accuracy. The trade-off is that higher-accuracy approximations generally require longer evolution times during the \textit{active phase}. We provided explicit formulae for the tuning parameters in this special set and examined representative examples, including the fastest-possible exact thermalization achievable in our scheme.

Beyond its conceptual simplicity, the protocol is broadly realizable: the required changes in coupling and frequency can be implemented in standard control operations in many experimental platforms, including, notably trapped ions\cite{Ospelkaus_2011}. Moreover, the scheme naturally connects to related quantum thermodynamic tasks such as refrigeration\cite{PhysRevA.97.032104}, state steering\cite{Kienzler_2015}, or finite-time implementation of isochoric strokes in quantum heat engines\cite{Ro_nagel_2014,Ro_nagel_2016}.

Looking ahead, this framework points toward a minimal setting for thermodynamic experiments, namely the realization of a “single-ion thermal universe,” where two transverse motional modes of a single ion effectively play the roles of system and bath. The coupled-oscillator model analyzed here provides a first step in this direction, as it captures the essential analytical structure of the problem and clarifies how the relevant control parameters must be tuned within an idealized setting.

However, extending this framework to a realistic setting requires addressing several nontrivial challenges, including the precise engineering of the required control protocols, as well as the impact of environmental coupling, additional noise sources, and experimental imperfections. We leave a detailed investigation of these aspects, including realistic numerical modeling and experimental feasibility, to future work.

Several promising directions remain for future work. First, a comprehensive numerical survey of the thermalization condition’s solution space could uncover faster or more robust protocols beyond the current \textit{special discrete set}. Second, incorporating realistic experimental constraints—such as finite quench rates, anharmonicity \cite{deffner2010quantum,zheng2014work}, and decoherence—would provide a clearer picture of practical performance. Third, generalizing the approach to multimode or non-Gaussian regimes could yield richer thermalization dynamics and broaden its applicability. In the multimode setting, potential links to the quantum marginal problem \cite{Klyachko_2006} and possible implications for Hawking radiation \cite{Aurell:2023wqa,Aurell:2025psi} would be especially compelling to explore.

In summary, our results show that finite-time, bath-free thermalization is not only theoretically possible but also analytically tractable in a large and experimentally relevant subset of cases.


\section*{Acknowledgments}
KR is supported by  EPSRC Standard Grant EP/X024199/1. The authors thank Sreenath K Manikandan and S Mahesh Chandran for several useful discussions. This work is part of MH’s master’s thesis; he sincerely thanks his internal guides, Shibi Thomas and Manesh Michael, for their comments and encouragement. MH also expresses gratitude to Lini Devassy and Anvy Moly Tom for their support and motivation.



\bibliographystyle{utphys1}
\bibliography{Heating_SHO}

@article{Chen:2023cuc,
    author = "Chen, Chi-Fang and Kastoryano, Michael J. and Brand{\~a}o, Fernando G. S. L. and Gily{\'e}n, Andr{\'a}s",
    title = "{Quantum Thermal State Preparation}",
    eprint = "2303.18224",
    archivePrefix = "arXiv",
    primaryClass = "quant-ph",
    month = "3",
    year = "2023"
}

@article{Vinjanampathy_2016,
   title={Quantum thermodynamics},
   volume={57},
   ISSN={1366-5812},
   url={http://dx.doi.org/10.1080/00107514.2016.1201896},
   DOI={10.1080/00107514.2016.1201896},
   number={4},
   journal={Contemporary Physics},
   publisher={Informa UK Limited},
   author={Vinjanampathy, Sai and Anders, Janet},
   year={2016},
   month=jul, pages={545–579} }

@article{Myers:2022lvm,
    author = "Myers, Nathan M. and Abah, Obinna and Deffner, Sebastian",
    title = "{Quantum thermodynamic devices: From theoretical proposals to experimental reality}",
    eprint = "2201.01740",
    archivePrefix = "arXiv",
    primaryClass = "quant-ph",
    reportNumber = "LA-UR-22-20089",
    doi = "10.1116/5.0083192",
    journal = "AVS Quantum Sci.",
    volume = "4",
    number = "2",
    pages = "027101",
    year = "2022"
}

@article{Klyachko_2006,
   title={Quantum marginal problem and N-representability},
   volume={36},
   ISSN={1742-6596},
   url={http://dx.doi.org/10.1088/1742-6596/36/1/014},
   DOI={10.1088/1742-6596/36/1/014},
   journal={Journal of Physics: Conference Series},
   publisher={IOP Publishing},
   author={Klyachko, Alexander A},
   year={2006},
   month=apr, pages={72–86} }

@article{Ro_nagel_2016,
   title={A single-atom heat engine},
   volume={352},
   ISSN={1095-9203},
   url={http://dx.doi.org/10.1126/science.aad6320},
   DOI={10.1126/science.aad6320},
   number={6283},
   journal={Science},
   publisher={American Association for the Advancement of Science (AAAS)},
   author={Roßnagel, Johannes and Dawkins, Samuel T. and Tolazzi, Karl N. and Abah, Obinna and Lutz, Eric and Schmidt-Kaler, Ferdinand and Singer, Kilian},
   year={2016},
   month=apr, pages={325–329} 
}

@article{Ro_nagel_2014,
   title={Nanoscale Heat Engine Beyond the Carnot Limit},
   volume={112},
   ISSN={1079-7114},
   url={http://dx.doi.org/10.1103/PhysRevLett.112.030602},
   DOI={10.1103/physrevlett.112.030602},
   number={3},
   journal={Physical Review Letters},
   publisher={American Physical Society (APS)},
   author={Roßnagel, J. and Abah, O. and Schmidt-Kaler, F. and Singer, K. and Lutz, E.},
   year={2014},
   month=jan }

@misc{deffner2019qtqi,
      title={Quantum Thermodynamics: An introduction to the thermodynamics of quantum information}, 
      author={Sebastian Deffner and Steve Campbell},
      year={2019},
      eprint={1907.01596},
      archivePrefix={arXiv},
      primaryClass={quant-ph},
      url={https://arxiv.org/abs/1907.01596}, 
}

@article{Goold2016role,
  author       = {John Goold and Arnau Riera and Lídia del Rio and Marcus Huber and Paul Skrzypczyk},
  title        = {The role of quantum information in thermodynamics -- a topical review},
  journal      = {Journal of Physics A: Mathematical and Theoretical},
  volume       = {49},
  number       = {14},
  pages        = {143001},
  year         = {2016},
  month        = feb,
  doi          = {10.1088/1751-8113/49/14/143001},
  publisher    = {IOP Publishing}
}

@article{Millen_Xuereb_2016,
  author       = {James Millen and Andr\'{e} Xuereb},
  title        = {Perspective on quantum thermodynamics},
  journal      = {New Journal of Physics},
  volume       = {18},
  number       = {1},
  pages        = {011002},
  year         = {2016},
  doi          = {10.1088/1367-2630/18/1/011002},
  publisher    = {IOP Publishing}
}

@article{Witten:2024upt,
    author = "Witten, Edward",
    title = "{Introduction to black hole thermodynamics}",
    eprint = "2412.16795",
    archivePrefix = "arXiv",
    primaryClass = "hep-th",
    doi = "10.1140/epjp/s13360-025-06288-y",
    journal = "Eur. Phys. J. Plus",
    volume = "140",
    number = "5",
    pages = "430",
    year = "2025"
}

@article{Harlow:2014yka,
    author = "Harlow, Daniel",
    title = "{Jerusalem Lectures on Black Holes and Quantum Information}",
    eprint = "1409.1231",
    archivePrefix = "arXiv",
    primaryClass = "hep-th",
    doi = "10.1103/RevModPhys.88.015002",
    journal = "Rev. Mod. Phys.",
    volume = "88",
    pages = "015002",
    year = "2016"
}

@article{Georgescu:2013oza,
    author = "Georgescu, I. M. and Ashhab, S. and Nori, Franco",
    title = "{Quantum Simulation}",
    eprint = "1308.6253",
    archivePrefix = "arXiv",
    primaryClass = "quant-ph",
    doi = "10.1103/RevModPhys.86.153",
    journal = "Rev. Mod. Phys.",
    volume = "86",
    pages = "153",
    year = "2014"
}

@article{Wineland:1997mg,
    author = "Wineland, D. J. and Monroe, C. and Itano, W. M. and Leibfried, D. and King, B. E. and Meekhof, D. M.",
    title = "{Experimental issues in coherent quantum-state manipulation of trapped atomic ions}",
    eprint = "quant-ph/9710025",
    archivePrefix = "arXiv",
    doi = "10.6028/jres.103.019",
    journal = "J. Res. Natl. Inst. Stand. Tech.",
    volume = "103",
    number = "3",
    pages = "259",
    year = "1998"
}

@article{Wallraff2004,
  author = {A. Wallraff and D. I. Schuster and A. Blais and L. Frunzio and R.-S. Huang and J. Majer and S. Kumar and S. M. Girvin and R. J. Schoelkopf},
  title = {Strong coupling of a single photon to a superconducting qubit using circuit quantum electrodynamics},
  journal = {Nature},
  volume = {431},
  number = {7005},
  pages = {162--167},
  year = {2004},
  doi = {10.1038/nature02851},
  url = {https://www.nature.com/articles/nature02851}
}

@article{Aspelmeyer:2013lha,
    author = "Aspelmeyer, Markus and Kippenberg, Tobias J. and Marquardt, Florian",
    title = "{Cavity Optomechanics}",
    eprint = "1303.0733",
    archivePrefix = "arXiv",
    primaryClass = "cond-mat.mes-hall",
    doi = "10.1103/RevModPhys.86.1391",
    journal = "Rev. Mod. Phys.",
    volume = "86",
    pages = "1391",
    year = "2014"
}

@article{Bachtold:2022hnr,
    author = "Bachtold, Adrian and Moser, Joel and Dykman, M. I.",
    title = "{Mesoscopic physics of nanomechanical systems}",
    eprint = "2202.01819",
    archivePrefix = "arXiv",
    primaryClass = "cond-mat.mes-hall",
    doi = "10.1103/RevModPhys.94.045005",
    journal = "Rev. Mod. Phys.",
    volume = "94",
    number = "4",
    pages = "045005",
    year = "2022"
}

@article{Torrontegui:2013sge,
    author = "Torrontegui, Erik and Ib{\'a}{\~n}ez, Sara and Mart{\'\i}nez-Garaot, Sofia and Modugno, Michele and del Campo, Adolfo and Gu{\'e}ry-Odelin, David and Ruschhaupt, Andreas and Chen, Xi and Muga, Juan Gonzalo",
    title = "{Shortcuts to Adiabaticity}",
    eprint = "1212.6343",
    archivePrefix = "arXiv",
    primaryClass = "quant-ph",
    reportNumber = "LA-UR-12-25725",
    doi = "10.1016/B978-0-12-408090-4.00002-5",
    journal = "Adv. At. Mol. Opt. Phys.",
    volume = "62",
    pages = "117--169",
    year = "2013"
}

@article{del_Campo_2019,
   title={Focus on Shortcuts to Adiabaticity},
   volume={21},
   ISSN={1367-2630},
   url={http://dx.doi.org/10.1088/1367-2630/ab1437},
   DOI={10.1088/1367-2630/ab1437},
   number={5},
   journal={New Journal of Physics},
   publisher={IOP Publishing},
   author={del Campo, Adolfo and Kim, Kihwan},
   year={2019},
   month=may, pages={050201} 
}

@article{ermakov1880,
  author  = {Ermakov, V. P.},
  title   = {Second order differential equations: conditions of complete integrability},
  journal = {Univ. Izv. Kiev},
  volume  = {20},
  number  = {9},
  pages   = {1},
  year    = {1880},
DOI={10.2298/AADM0802123E},
}

@article{lewis1967classical,
  title={Classical and quantum systems with time-dependent harmonic-oscillator-type Hamiltonians},
  author={Lewis Jr, H Ralph},
  journal={Physical Review Letters},
  volume={18},
  number={13},
  pages={510},
  year={1967},
  publisher={APS},
DOI={10.1103/PhysRevLett.18.510}
}

@article{lewis1968class,
  title={Class of exact invariants for classical and quantum time-dependent harmonic oscillators},
  author={Lewis Jr, H Ralph},
  journal={Journal of Mathematical Physics},
  volume={9},
  number={11},
  pages={1976--1986},
  year={1968},
  publisher={American Institute of Physics},
DOI={10.1063/1.1664532}
}

@article{monroe1995resolved,
  title={Resolved-sideband Raman cooling of a bound atom to the 3D zero-point energy},
  author={Monroe, Ch and Meekhof, DM and King, BE and Jefferts, Steven R and Itano, Wayne M and Wineland, David J and Gould, P},
  journal={Physical review letters},
  volume={75},
  number={22},
  pages={4011},
  year={1995},
  publisher={APS},
DOI={ 10.1103/PhysRevLett.75.4011}
}

@article{meekhof1996generation,
  title={Generation of nonclassical motional states of a trapped atom},
  author={Meekhof, DM and Monroe, C and King, BE and Itano, Wayne M and Wineland, David J},
  journal={Physical review letters},
  volume={76},
  number={11},
  pages={1796},
  year={1996},
  publisher={APS},
DOI={ 10.1103/PhysRevLett.76.1796}
}

@article{deffner2008nonequilibrium,
  title={Nonequilibrium work distribution of a quantum harmonic oscillator},
  author={Deffner, Sebastian and Lutz, Eric},
  journal={Physical Review E—Statistical, Nonlinear, and Soft Matter Physics},
  volume={77},
DOI={10.1103/PhysRevE.77.021128},
  number={2},
  pages={021128},
  year={2008},
  publisher={APS}
}

@book{kirk2004optimal,
  title        = {Optimal Control Theory: An Introduction},
  author       = {Kirk, Donald E.},
  year         = {2004},
  publisher    = {Courier Corporation},
  address      = {Mineola, NY},
  isbn         = {9780486434841},
  url          = {https://books.google.com/books/about/Optimal_Control_Theory.html?id=fCh2SAtWIdwC}
}

@article{zheng2014work,
  title={Work and efficiency of quantum Otto cycles in power-law trapping potentials},
  author={Zheng, Yuanjian and Poletti, Dario},
  journal={Physical Review E},
  volume={90},
  number={1},
  pages={012145},
  year={2014},
  publisher={APS},
DOI={10.1103/PhysRevE.90.012145}
}

@article{deffner2010quantum,
  title={Quantum work statistics of linear and nonlinear parametric oscillators},
  author={Deffner, Sebastian and Abah, Obinna and Lutz, Eric},
  journal={Chemical Physics},
  volume={375},
  number={2-3},
  pages={200--208},
  year={2010},
  publisher={Elsevier},
DOI={10.1016/j.chemphys.2010.04.042}
}

@article{Horodecki_2013,
   title={Fundamental limitations for quantum and nanoscale thermodynamics},
   volume={4},
   ISSN={2041-1723},
   url={http://dx.doi.org/10.1038/ncomms3059},
   DOI={10.1038/ncomms3059},
   number={1},
   journal={Nature Communications},
   publisher={Springer Science and Business Media LLC},
   author={Horodecki, Michał and Oppenheim, Jonathan},
   year={2013},
   month=jun }

@article{H_nggi_2009,
   title={Artificial Brownian motors: Controlling transport on the nanoscale},
   volume={81},
   ISSN={1539-0756},
   url={http://dx.doi.org/10.1103/RevModPhys.81.387},
   DOI={10.1103/revmodphys.81.387},
   number={1},
   journal={Reviews of Modern Physics},
   publisher={American Physical Society (APS)},
   author={Hänggi, Peter and Marchesoni, Fabio},
   year={2009},
   month=mar, pages={387–442} 
}

@article{Kosloff_2017,
   title={The Quantum Harmonic Otto Cycle},
   volume={19},
   ISSN={1099-4300},
   url={http://dx.doi.org/10.3390/e19040136},
   DOI={10.3390/e19040136},
   number={4},
   journal={Entropy},
   publisher={MDPI AG},
   author={Kosloff, Ronnie and Rezek, Yair},
   year={2017},
   month=mar, pages={136}
}

@article{Deffner_2018,
   title={Efficiency of Harmonic Quantum Otto Engines at Maximal Power},
   volume={20},
   ISSN={1099-4300},
   url={http://dx.doi.org/10.3390/e20110875},
   DOI={10.3390/e20110875},
   number={11},
   journal={Entropy},
   publisher={MDPI AG},
   author={Deffner, Sebastian},
   year={2018},
   month=nov, pages={875} 
}

@article{Serafini_2020,
   title={Gaussian Thermal Operations and The Limits of Algorithmic Cooling},
   volume={124},
   ISSN={1079-7114},
   url={http://dx.doi.org/10.1103/PhysRevLett.124.010602},
   DOI={10.1103/physrevlett.124.010602},
   number={1},
   journal={Physical Review Letters},
   publisher={American Physical Society (APS)},
   author={Serafini, A. and Lostaglio, M. and Longden, S. and Shackerley-Bennett, U. and Hsieh, C.-Y. and Adesso, G.},
   year={2020},
   month=jan }

@article{Singh_2020,
   title={Performance bounds of nonadiabatic quantum harmonic Otto engine and refrigerator under a squeezed thermal reservoir},
   volume={102},
   ISSN={2470-0053},
   url={http://dx.doi.org/10.1103/PhysRevE.102.062123},
   DOI={10.1103/physreve.102.062123},
   number={6},
   journal={Physical Review E},
   publisher={American Physical Society (APS)},
   author={Singh, Varinder and M{\"u}stecapl{\i}o{\u{g}}lu, {\"O}zg{\"u}r E},
   year={2020},
   month=dec 
}

@article{PhysRevA.91.020502,
  title = {Dynamical Casimir effect and minimal temperature in quantum thermodynamics},
  author = {Benenti, Giuliano and Strini, Giuliano},
  journal = {Phys. Rev. A},
  volume = {91},
  issue = {2},
  pages = {020502},
  numpages = {4},
  year = {2015},
  month = {Feb},
  publisher = {American Physical Society},
  doi = {10.1103/PhysRevA.91.020502},
  url = {https://link.aps.org/doi/10.1103/PhysRevA.91.020502}
}

@article{PhysRevE.93.062106,
  title = {Quantum jump model for a system with a finite-size environment},
  author = {Suomela, S. and Kutvonen, A. and Ala-Nissila, T.},
  journal = {Phys. Rev. E},
  volume = {93},
  issue = {6},
  pages = {062106},
  numpages = {7},
  year = {2016},
  month = {Jun},
  publisher = {American Physical Society},
  doi = {10.1103/PhysRevE.93.062106},
  url = {https://link.aps.org/doi/10.1103/PhysRevE.93.062106}
}

@article{Reid_2017,
   title={A self-contained quantum harmonic engine},
   volume={120},
   ISSN={1286-4854},
   url={http://dx.doi.org/10.1209/0295-5075/120/60006},
   DOI={10.1209/0295-5075/120/60006},
   number={6},
   journal={EPL (Europhysics Letters)},
   publisher={IOP Publishing},
   author={Reid, B. and Pigeon, S. and Antezza, M. and De Chiara, G.},
   year={2017},
   month=dec, pages={60006} 
}

@article{Ospelkaus_2011,
   title={Microwave quantum logic gates for trapped ions},
   volume={476},
   ISSN={1476-4687},
   url={http://dx.doi.org/10.1038/nature10290},
   DOI={10.1038/nature10290},
   number={7359},
   journal={Nature},
   publisher={Springer Science and Business Media LLC},
   author={Ospelkaus, C. and Warring, U. and Colombe, Y. and Brown, K. R. and Amini, J. M. and Leibfried, D. and Wineland, D. J.},
   year={2011},
   month=aug, pages={181–184} 
}

@article{Bera_2019,
   title={Thermodynamics as a Consequence of Information Conservation},
   volume={3},
   ISSN={2521-327X},
   url={http://dx.doi.org/10.22331/q-2019-02-14-121},
   DOI={10.22331/q-2019-02-14-121},
   journal={Quantum},
   publisher={Verein zur Forderung des Open Access Publizierens in den Quantenwissenschaften},
   author={Bera, Manabendra Nath and Riera, Arnau and Lewenstein, Maciej and Khanian, Zahra Baghali and Winter, Andreas},
   year={2019},
   month=feb, pages={121} 
}

@article{PhysRevA.97.032104,
  title = {Cooling a quantum oscillator: A useful analogy to understand laser cooling as a thermodynamical process},
  author = {Freitas, Nahuel and Paz, Juan Pablo},
  journal = {Phys. Rev. A},
  volume = {97},
  issue = {3},
  pages = {032104},
  numpages = {12},
  year = {2018},
  month = {Mar},
  publisher = {American Physical Society},
  doi = {10.1103/PhysRevA.97.032104},
  url = {https://link.aps.org/doi/10.1103/PhysRevA.97.032104}
}

@article{Kienzler_2015,
   title={Quantum harmonic oscillator state synthesis by reservoir engineering},
   volume={347},
   ISSN={1095-9203},
   url={http://dx.doi.org/10.1126/science.1261033},
   DOI={10.1126/science.1261033},
   number={6217},
   journal={Science},
   publisher={American Association for the Advancement of Science (AAAS)},
   author={Kienzler, D. and Lo, H.-Y. and Keitch, B. and de Clercq, L. and Leupold, F. and Lindenfelser, F. and Marinelli, M. and Negnevitsky, V. and Home, J. P.},
   year={2015},
   month=jan, pages={53–56} 
}

@article{Mancini:2006pju,
    author = "Mancini, Stefano and Wiseman, Howard M.",
    title = "{Optimal control of entanglement via quantum feedback}",
    eprint = "quant-ph/0610006",
    archivePrefix = "arXiv",
    doi = "10.1103/PhysRevA.75.012330",
    journal = "Phys. Rev. A",
    volume = "75",
    pages = "012330",
    year = "2007"
}

@article{Jussiau_2021,
   title={Thermal control across a chain of electronic nanocavities},
   volume={104},
   ISSN={2469-9969},
   url={http://dx.doi.org/10.1103/PhysRevB.104.045414},
   DOI={10.1103/physrevb.104.045414},
   number={4},
   journal={Physical Review B},
   publisher={American Physical Society (APS)},
   author= {Jussiau, \'Etienne and Manikandan, Sreenath K. and Bhandari, Bibek and Jordan, Andrew N.},
   year={2021},
   month=jul 
}

@article{Aurell:2025psi,
    author = "Aurell, Erik and Hackl, Lucas and Kieburg, Mario",
    title = "{Average entanglement entropy of a small subsystem in a constrained pure Gaussian state ensemble}",
    eprint = "2505.03696",
    archivePrefix = "arXiv",
    primaryClass = "quant-ph",
    month = "5",
    year = "2025"
}

@article{Aurell:2023wqa,
    author = "Aurell, Erik and Hackl, Lucas and Horodecki, Pawel and Jonsson, Robert H. and Kieburg, Mario",
    title = "{Random Pure Gaussian States and Hawking Radiation}",
    eprint = "2311.10562",
    archivePrefix = "arXiv",
    primaryClass = "gr-qc",
    reportNumber = "NORDITA 2023-094 (for co-author RHJ)",
    doi = "10.1103/PhysRevLett.133.060202",
    journal = "Phys. Rev. Lett.",
    volume = "133",
    number = "6",
    pages = "060202",
    year = "2024"
}

@misc{ding2025therm,
      title={End-to-End Efficient Quantum Thermal and Ground State Preparation Made Simple}, 
      author={Zhiyan Ding and Yongtao Zhan and John Preskill and Lin Lin},
      year={2025},
      eprint={2508.05703},
      archivePrefix={arXiv},
      primaryClass={quant-ph},
      url={https://arxiv.org/abs/2508.05703}, 
}

@article{Dupays:2020sta,
    author = "Dupays, Léonce and Egusquiza, Iñigo L. and del Campo, Adolfo and Chenu, Aurélia",
    title = "{Superadiabatic thermalization of a quantum oscillator by engineered dephasing}",
    doi = "10.1103/PhysRevResearch.2.033178",
    journal = "Phys. Rev. Research",
    volume = "2",
    pages = "033178",
    year = "2020"
}

@article{Alipour:2020sad,
    author = "Alipour, Sahar and Chenu, Aurélia and Rezakhani, Ali T. and del Campo, Adolfo",
    title = "{Shortcuts to Adiabaticity in Driven Open Quantum Systems}",
    doi = "10.22331/q-2020-09-28-336",
    journal = "Quantum",
    volume = "4",
    pages = "336",
    year = "2020"
}

@article{Santos:2024sta,
    author = "Santos, Alan C.",
    title = "{Shortcut-to-adiabaticity for coupled harmonic oscillators}",
    doi = "10.1140/epjp/s13360-024-05718-7",
    journal = "Eur. Phys. J. Plus",
    volume = "139",
    pages = "909",
    year = "2024"
}

@article{Poyatos:1996qre,
    author = "Poyatos, J. F. and Cirac, J. I. and Zoller, P.",
    title = "{Quantum Reservoir Engineering with Laser Cooled Trapped Ions}",
    doi = "10.1103/PhysRevLett.77.4728",
    journal = "Phys. Rev. Lett.",
    volume = "77",
    pages = "4728",
    year = "1996"
}

@article{So:2025etr,
    author = "So, Vanessa and others",
    title = "{Experimental Realization of Thermal Reservoirs with Tunable Temperature in a Trapped-Ion Spin-Boson Simulator}",
    eprint = "2511.08689",
    archivePrefix = "arXiv",
    primaryClass = "quant-ph",
    year = "2025"
}

@article{Metcalf:2020etc,
    author = "Metcalf, Mekena and Moussa, Jonathan E. and de Jong, Wibe A. and Sarovar, Mohan",
    title = "{Engineered thermalization and cooling of quantum many-body systems}",
    doi = "10.1103/PhysRevResearch.2.023214",
    journal = "Phys. Rev. Research",
    volume = "2",
    pages = "023214",
    year = "2020"
}

@article{Leibfried2003,
    author    = "Leibfried, D. and Blatt, R. and Monroe, C. and Wineland, D.",
    title     = "{Quantum dynamics of single trapped ions}",
    journal   = "Rev. Mod. Phys.",
    volume    = "75",
    pages     = "281",
    year      = "2003",
    doi       = "10.1103/RevModPhys.75.281"
}

@article{Ding:2017qpo,
    author    = "Ding, Shiqian and Maslennikov, Gleb and Habl{\"u}tzel, Roland 
                 and Loh, Huanqian and Matsukevich, Dzmitry",
    title     = "{Quantum Parametric Oscillator with Trapped Ions}",
    journal   = "Phys. Rev. Lett.",
    volume    = "119",
    number    = "15",
    pages     = "150404",
    year      = "2017",
    doi       = "10.1103/PhysRevLett.119.150404",
    eprint    = "1512.01670",
    archivePrefix = "arXiv",
    primaryClass  = "quant-ph"
}

@article{Diedrich1989,
    author    = "Diedrich, F. and Bergquist, J. C. and Itano, Wayne M. 
                 and Wineland, D. J.",
    title     = "{Laser Cooling to the Zero-Point Energy of Motion}",
    journal   = "Phys. Rev. Lett.",
    volume    = "62",
    pages     = "403",
    year      = "1989",
    doi       = "10.1103/PhysRevLett.62.403"
}

@article{Turchette2000,
    author    = "Turchette, Q. A. and Kielpinski, D. and King, B. E. and Leibfried, D. 
                 and Meekhof, D. M. and Myatt, C. J. and Rowe, M. A. and Sackett, C. A. 
                 and Wood, C. S. and Itano, W. M. and Monroe, C. and Wineland, D. J.",
    title     = "{Heating of trapped ions from the quantum ground state}",
    journal   = "Phys. Rev. A",
    volume    = "61",
    pages     = "063418",
    year      = "2000",
    doi       = "10.1103/PhysRevA.61.063418"
}

@article{Brownnutt2015,
    author    = "Brownnutt, M. and Kumph, M. and Rabl, P. and Blatt, R.",
    title     = "{Ion-trap measurements of electric-field noise near surfaces}",
    journal   = "Rev. Mod. Phys.",
    volume    = "87",
    pages     = "1419",
    year      = "2015",
    doi       = "10.1103/RevModPhys.87.1419"
}

@article{Jefferts1995,
    author    = "Jefferts, S. R. and Monroe, C. and Barton, A. S. and Wineland, D. J.",
    title     = "{Paul Trap for Optical Frequency Standards}",
    journal   = "IEEE Trans. Instrum. Meas.",
    volume    = "44",
    number    = "2",
    pages     = "148--150",
    year      = "1995",
    doi       = "10.1109/19.377786"
}

\end{document}